\documentclass[reqno,12pt]{olplainarticle}

\usepackage{threeparttable}
\usepackage{siunitx}
\usepackage{mathtools}

\newtagform{brackets}{[}{]}
\usetagform{brackets}

\DeclareMathAlphabet{\mathcal}{OMS}{cmsy}{m}{n}
\newcommand\blfootnote[1]{%
  \begingroup
  \renewcommand\thefootnote{}\footnote{#1}%
  \addtocounter{footnote}{-1}%
  \endgroup
}

\title{Chiral interactions between tropocollagen molecules determine the collagen microfibril structure}

\author[a]{Art'om Zolotarjov}
\author[b,2]{Roland Kr\"oger}
\author[a,3]{Dmitri O. Pushkin}
\affil[a]{Department of Mathematics, University of York, York}
\affil[b]{School of PET, University of York, York}


\begin{abstract}
Collagen is the most abundant structural protein in animals, forming hierarchically organised fibrils that provide mechanical support to tissues. Despite detailed structural studies, the physical principles that govern the formation of the characteristic D-periodic collagen microfibril remain poorly understood. Here, we present a theoretical framework that links the amino acid sequence of tropocollagen to its supramolecular organisation. By combining statistical modeling of residue geometry with sequence-informed interaction potentials, we show that the chiral arrangement of outward-facing residues induces directional intermolecular interactions that drive molecular supercoiling. These interactions favour the formation of right-handed, pentameric microfibrils with a staggered axial periodicity of approximately 67 nm. Our simulations reveal that this structure emerges across a wide range of mammalian collagen sequences as a global energy minimum robust to biochemical noise. These findings provide a mechanistic explanation for collagen’s supramolecular chirality and offer design principles for engineering synthetic collagen-mimetic materials.
\end{abstract}

\begin{document}
\flushbottom
\maketitle
\thispagestyle{empty}

\section*{Introduction}
\blfootnote{Correspondence E-mail: $^{3}$mitya.pushkin@york.ac.uk and
$^{2}$roland.kroger@york.ac.uk}

Collagen is by far the most abundant protein in the extracellular matrix,  connective tissues, skin, and bones \cite{bella2017fibrillar,reznikov2018fractal}. It provides the scaffold that enables the organisation of cells into tissues. As a key structural biomaterial, it influences a multitude of multicellular processes, from bone mineralization to invasions of cancer cells \cite{reznikov2018fractal,koorman2022spatial} and has even been linked to the Cambrian explosion of multicellular life \cite{towe1970oxygen,fidler2017collagen}. Since collagen is essential for maintaining tissue structure and function, it is a key focus in regenerative medicine, wound healing, orthopedics, dermatology, and cardiovascular health.
Recent advances in biochemical engineering have produced the amino acid sequences of thousands of natural collagens \cite{o2016reference} and have
enabled the design of synthetic collagen mimetic peptides (CMPs)\cite{yu2011collagen}. The biocompatibility of CMPs, their tunable properties, and their potential to replicate natural collagen structures make them indispensable for tissue engineering \cite{xu2021collagen}. 

At the physical level, the versatility of collagen as a structural protein arises from its propensity to assemble into fibrils, bundles of fibrils, and intricate hierarchical fibrillar matrices. Despite its importance, the physical principles of collagen self-assembly and their link to the amino acid sequences remain poorly understood.  
Tropocollagen, the smallest unit in the fibrillar hierarchy, is a semiflexible molecule approximately $L \approx 300 \, \text{nm} $ long \cite{rezaei2018environmentally} and $1.5 \, \text{nm}$ in diameter. It
is made up of three polypeptide strands ($\alpha$ chains) wrapped around in a right-handed helix, see Fig. \ref{StructureFig}(A). In each strand, about $ 1000$ amino acids are arranged in a sequence with the regular motif of [Gly-X-Y], where Gly is glycine, and X and Y may be various amino acids but most often proline (Pro) and hydroxyproline (Hyp), respectively. Tropocollagen is classified into nearly thirty distinct types, each varying
in amino acid composition and the hierarchical structures they form \cite{bella2017fibrillar}. In this work, we focus on the structures formed by fibrillar collagens, with type I collagen being the most abundant. It has been extensively studied in experimental literature \cite{orgel2006microfibrillar}. 

\begin{figure*}[!t]
\centering
\includegraphics[width=17.8 cm, trim ={0cm 0cm 0cm 0cm}, clip, width = \linewidth]{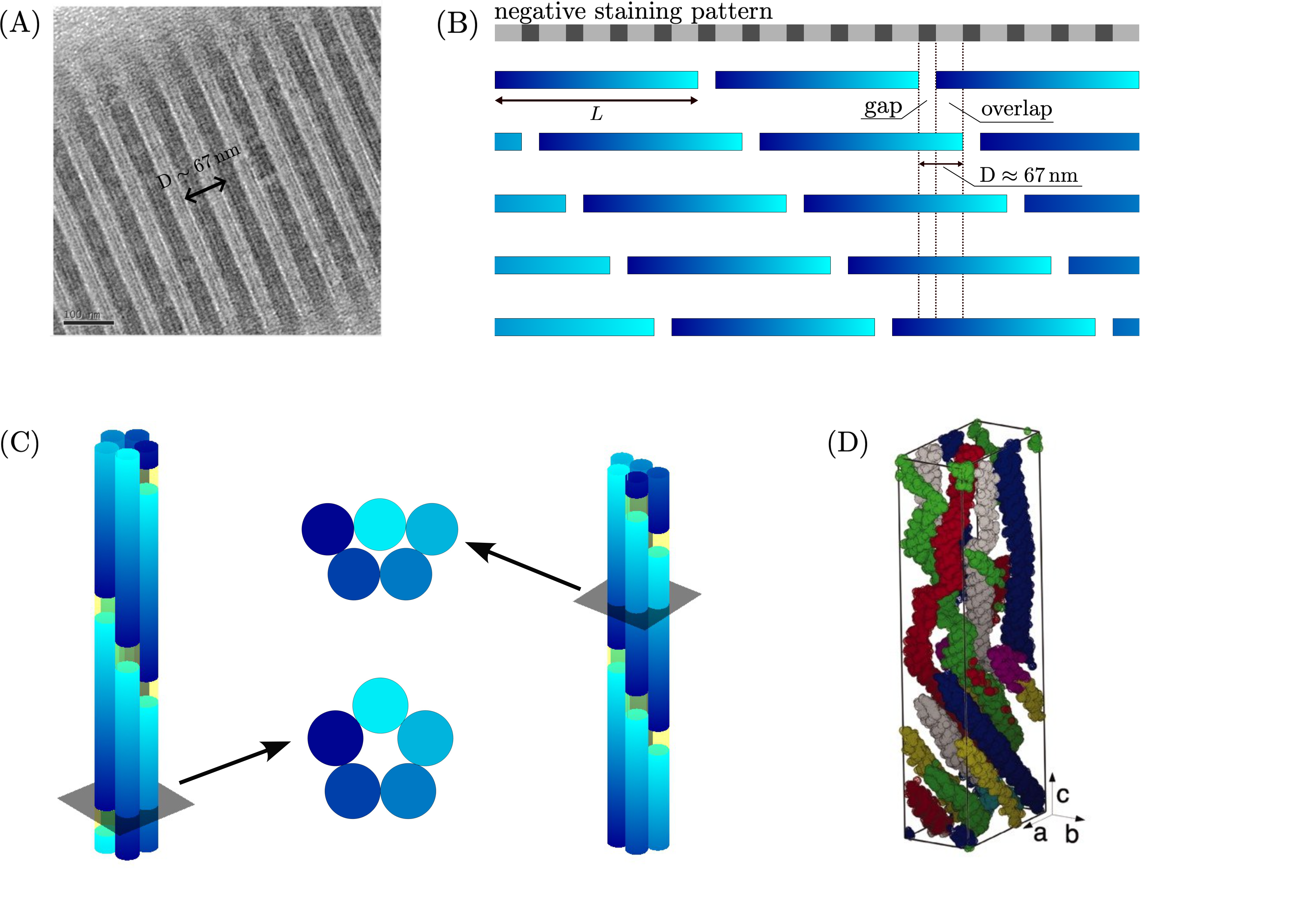}
\caption{(A). TEM image of a negatively stained collagen fibril showing the D-banding pattern (image obtained from \cite{wieczorek2015development}). (B). 2D representation of a collagen microfibril according to the Hodge-Petruska scheme. (C). Schematic representations of the 3D microfibril models. Gap regions are highlighted in yellow. (LEFT) Smith microfibril. (RIGHT) Compressed microfibril. (D). Orgel model of the collagen unit cell (obtained from \cite{orgel2006microfibrillar}).}
\label{HistOverview}
\end{figure*}

Tropocollagen readily assembles
{\em in vivo} and {\em in vitro} in fibrils, with typical diameters between $10 \, \text{nm}$ and $100 \, \text{nm}$ \cite{revell2021collagen}. Their length is orders of magnitude larger than their diameters. 
One of their salient features is the periodic axial density modulations, which appear as regular alternating light and dark bands with period $\text{D} \approx 67 \text{nm}$ in negatively stained TEM samples, see Fig. \ref{HistOverview}A. D is often called the native collagen period. The period value, $\text{D}$, is highly conserved across different collagen types, although notable variations occasionally occur, even within the same tissue \cite{fang2012type,bella2017fibrillar}. The regularity of the banding pattern is crucial for the mechanical strength of fibrils and collagen-rich tissues \cite{chen2017relationship}.

A simplified and widely used explanation for the banding pattern, known as the Hodge-Petruska scheme, was proposed in 1964 (see Fig. \ref{HistOverview}B) \cite{petruska1964subunit}. It envisages a two-dimensional stack of aligned straight molecules of length $L$, each shifted by the distance $\text{D}$ relative to its neighbour. The banding pattern is explained by the assumption that fibrils are composed of pentameric units. Since $L \approx 4.46 \text{D}$, five tropocollagen molecules staggered according to the Hodge-Petruska scheme create alternating `overlap' regions of length $0.46 \text{D}$ that contain five molecules and `gap' regions of length $0.54 \text{D}$ that contain four molecules. The later Smith's microfibril model expanded on this scheme by positioning the five neighbouring molecules at the vertices of a regular pentagon in a plane normal to the fibrillar axis (see Fig. \ref{HistOverview}C) \cite{smith1968molecular}. To further reconcile this model with the quasi-hexagonal lateral packing of individual tropocollagen molecules observed in experiments, the compressed Smith's microfibril model was proposed (see Fig. \ref{HistOverview}C) \cite{trus1980compressed}. 

Since 1964, the reality of microfibrils and their role as basic blocks of the fibril, have been experimentally confirmed \cite{orgel2001situ,orgel2006microfibrillar}. In a series of papers, Orgel and co-workers have resolved {\em in situ} the molecular structure of the microfibril using multiple isomorphous replacement and X-ray diffraction experiments. In particular, they showed that five neighbouring molecules are arranged to form a supertwisted right-handed microfibril that interdigitates with neighbouring microfibrils, see Fig. \ref{HistOverview}D. This interdigitation establishes the crystallographic superlattice, which is formed of quasi-hexagonally packed collagen molecules. 

Despite this progress, the theoretical foundations of these packing schemes remain unclear. What physical interactions result in a pentameric microfibril? What determines the axial stagger distance $\text{D}$? What governs the handedness of the microfibril? And how is this information encoded within the amino acid sequence? While some of these questions have occasionally been addressed in theoretical studies \cite{hulmes1973analysis,trus1976molecular,piez1978sequence,hofmann1978role,chen1991energetic,puszkarska2022using}, to the best of the authors' knowledge, the microfibrillar structure has not been explained as an emergent phenomenon arising from the fundamental molecular interactions. 

Orthologous $\alpha$ chain sequences have been extensively documented for all known fibrillar collagens \cite{o2016reference}. An important model predicting the axial stagger between pairs of tropocollagen molecules comprising the microfibril using residue sequence data was recently suggested by Puszkarska {\em et al.} \cite{puszkarska2022using}. In this approach, the D-stagger emerges as the equilibrium microfibril configuration corresponding to a local minimum of the free energy. The interaction potential between pairs of collagen molecules is calculated using the Miyazawa-Jernigan approximation for the contact interaction energy between amino acids, averaged over all possible inter-residue contacts.
The latter were determined based on the spatial proximity of residues, which is directly related to the sequence proximity: two residues close in a sequence are necessarily close in space. Hence, one can think of this algorithm as calculating interactions between linear sequences of residues.  Consequently, this algorithm ignores the angular dependence of the interaction potential between two collagen molecules. This drawback, in particular, precludes explaining the emergent supertwist and handedness of the microfibril.  
  
In this article, we extend the approach of \cite{puszkarska2022using}, combining empirical studies with theoretical arguments to quantify the interaction potential between pairs of parallel tropocollagen molecules. Using numerical simulations, we investigate which features of this potential, and under what conditions, give rise to the microfibrillar structure. In particular, we demonstrate that the pairwise interactions between collagen molecules are chiral due to the chiral spatial arrangement of the outward-facing residues of tropocollagen. This chirality is propagated to the level of the microfibrillar aggregate,  resulting in the right-handed supertwist of individual tropocollagen molecules. Furthermore, we attribute the pentameric nature of the microfibril to the geometric arrangement of residues. Our findings reveal that the optimal axial stagger, $\Delta z$, can assume different values corresponding to distinct local free energy minima, most notably $\Delta z \approx 0$ and $\Delta z = n \text{D}$, where $n =1,\dots,4$. The local minimum at $\Delta z \approx 0$ has until now been largely overlooked in theoretical discussions of microfibrillar structures despite experimental evidence for the existence of segment-long spacing (SLS) aggregates both \textit{in vitro} and \textit{in vivo} \cite{harris2016collagen,hulmes1983state}. We show that while the minimum at $\Delta z \approx 0$ is generally stronger than those at $\Delta z = n \text{D}$, it is sensitive to noise in the residue-residue interaction energies. Consequently, the axial staggers $\Delta z = n\text{D}$ emerge as robust global optima under most conditions. 

To compare our predictions with the available experimental data, we analyze amino acid sequences of more than $1000$ known fibril-forming collagens of mammalian species. In the absence of detailed studies of the microfibrillar structure for most of the collagens, we take the experimentally observed $\text{D}$-banding pattern in macroscopic aggregates as a proxy for the formation of the D-staggered microfibril. Under this assumption, we predict that all 176 analyzed sequences of heterotrimeric type I collagen result in D-banding, in agreement with the general knowledge in the field \cite{stylianou2022assessing}. This agreement validates our methods and lends credibility to our predictions for other, less well-studied collagen types.

\section*{Results}
\subsection*{Chiral Interactions and Helical Strip Organisation in Tropocollagen}

\begin{figure*}[!t]
\centering
\makebox[\textwidth][c]{\includegraphics[scale=0.6]{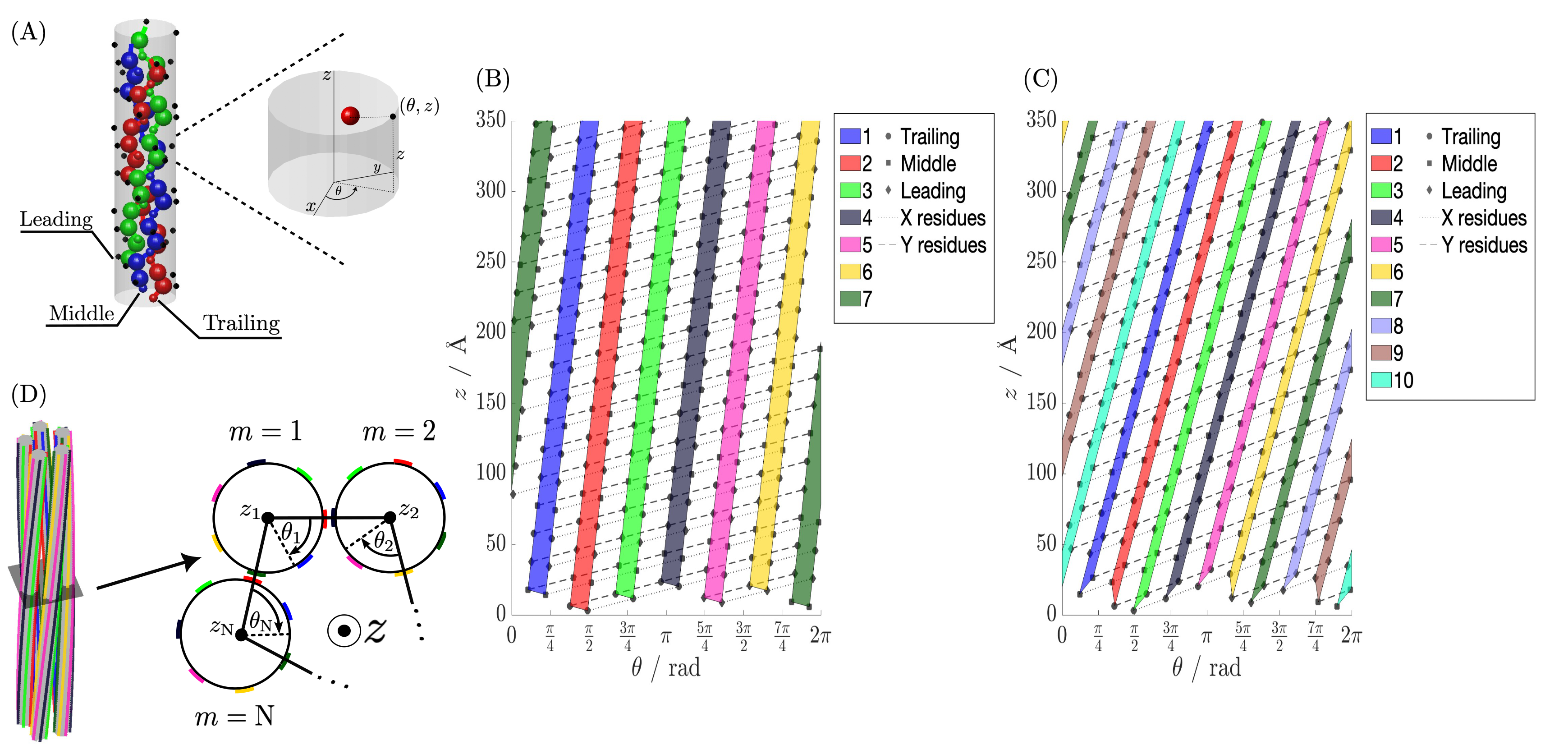}}%
\caption{(A). Segment of the collagen triple helix bounded by a cylindrical surface onto which coordinates of each $\text{C}_{\alpha}$ residue atom are projected. The $\text{C}_{\alpha}$-atom positions are obtained using two statistically derived parametrisations based on analysis of (B) Pro-rich and (C) Pro-poor model peptides. Dotted lines connect the residues that belong to the same $\alpha$ chain. We conventionally denote the most N-terminal $\alpha$ chain as trailing. Solid lines indicate imaginary connections between residues that fall on the same spiral strip. The spiral strips are numbered in order of appearance when moving counter-clockwise around the molecular $z$-axis and the most N-terminal residue is assigned the azimuthal coordinate $\theta=\frac{\pi}{2}$. The Pro-rich and Pro-poor parametrisations give rise to the two families of right-handed helical strips of amino acids with $7$ and $10$ helices in each family, respectively. (D). N-membered collagen microfibril model. An axially periodic microfibril is comprised of aligned tropocollagen
molecules placed at the vertices of a regular N-gon in the
azimuthal plane. The coloured segments on the molecular surfaces correspond to the Pro-rich strips shown in (B).}
\label{StructureFig}
\end{figure*}

The observed molecular supertwist of tropocollagen molecules within the microfibril points to the chiral nature of intermolecular interactions. We trace this chirality to the spatial arrangement of the outward-facing residues of the tropocollagen molecule. 

High-resolution data on the spatial organisation of collagen residues is currently unavailable, and several distinct structural models attempt to describe it on average \cite{bella2010new}. To avoid choosing between the models, we use a statistically-derived parametrisation of the triple helix based on the analysis of multiple high-resolution structures of shorter peptides modeling sections of the triple helix \cite{rainey2002statistically}. This statistical model accounts for differences in the imino acid content, resulting in two distinct triple helix parameterizations: Pro-rich and Pro-poor.  These parametrisations can also be viewed as limiting cases, representing the helical parameters of the average collagen structure corresponding to triple-helix segments that are entirely saturated or completely free of Pro and/or Hyp residues \cite{bella2010new}. 

We now demonstrate that each of the parametrisations gives rise to a helical arrangement of the outward-facing residues. Fig. \ref{StructureFig}A shows the positions of the residues projected on the cylindrical surface of a molecule unwrapped on the $(\theta,z)$-plane, where $\theta$ represents the azimuthal angle and $z$ is the axial position for each parametrisation. The position of each residue is indicated by the location of its $\text{C}_{\alpha}$ atom. The outward-facing residues cluster into two families of equally spaced, right-handed helical strips, as shown in Figures \ref{StructureFig}B and \ref{StructureFig}C. For the Pro-rich parametrisation, there are seven helical strips, each separated azimuthally by $2 \pi/7 \,\text{rad} \approx 51.4^\circ$, with a pitch of approximately $\SI[]{200}{\nano\metre}$.  For the Pro-poor parametrisation, there are ten helical strips, with an azimuthal separation of $2\pi/10 \, \text{rad}= 36^\circ$ and a pitch of approximately $ \SI[]{75} {\nano\metre}$. These emergent helical strips are distinct from the helices formed by the sequential positions of residues.
The helical strips have a finite width of approximately \SI[]{21}{\degree} for the Pro-rich case and \SI[]{16}{\degree} for the Pro-poor case. This width arises from the constant azimuthal coordinate difference between the left and right edges of each strip, which are uniformly composed of X and Y residues, respectively. It is important to note that within a given strip, the spatially nearest X and Y residues do not belong to the same [Gly-X-Y] triplet.

The actual spatial distribution of outward-facing residues on
the surface of the tropocollagen molecule varies as a function
of the amino acid composition \cite{orgel2014variation}. It is, clearly, chiral (non-superimposable with its mirror image). This chirality generates a chiral interaction potential between pairs of parallel tropocollagen molecules, leading to torques that can bend and twist them.
While the potential may exhibit complex dependencies on the relative orientation of the molecules, we show that only certain features of this potential are essential for forming the axially periodic, helical microfibril. This enables the development of a simplified model sufficient to predict the microfibril structure. We assume that the outward-facing residues of a tropocollagen molecule can be represented as a superposition of helical strip families with varying pitches \cite{rainey2002statistically, orgel2014variation}, such as the Pro-rich and Pro-poor families described in \cite{rainey2002statistically}. When two parallel molecules are close enough for their outward-facing residues to interact, strong interactions between the residues in their helical strips generate torques that bend and twist the molecules to minimize interaction energy, aligning the strips along a common axis. This process is analogous to molecular supercoiling observed in coiled coils, which arises from chiral interactions between hydrophobic strips on $\alpha$-helices \cite{neukirch2008chirality, liu2004atomic}, though collagen differs in having multiple helical strip families. For the interaction to dominate, the energy gained must outweigh the bending energy cost. This condition is easily met for the 7-strip family but is highly restrictive for the 10-strip family due to their smaller pitch and a strong dependence of the elastic deformation cost on the pitch (see Materials and Methods). Consequently, interactions from the 7-strip family are energetically favoured.

Thus, we model the effective interaction potential between collagen molecules based on the 7 helical strips from the Pro-rich parameterization. This leads to the prediction that collagen molecules in a microfibril form right-handed helices with a helical angle of approximately 
$5^\circ$, see Eq. [\ref{e:phiStar}] in Materials and Methods, consistent with experimental observations in bone and tendon \cite{baselt1993subfibrillar, raspanti2018not}.

\subsection*{Energetics of Strip-Strip Interactions}
Based on this, we hypothesize that interactions between aligned collagen molecules in a microfibril are primarily driven by opposing strip-strip interactions. With 7 strips, there are 28 potential strip-strip interactions, denoted as $\text{E}_{i\text{-}j}(\Delta z)$, where $ -L \le \Delta z \le L$ is the axial stagger of strip $j$ relative to strip $i$. We calculate them using the empirically determined Miyazawa-Jernigan  contact potentials (MJCP) for residue-residue interactions, see Materials and Methods. 

Assuming that collagen molecules form axially periodic arrays separated by gaps of length $g$, the interactions between opposing arrays are described by 28 $(L+g)$-periodic potentials: $\text{E}^{p}_{i\text{-}j}(\Delta z) = \text{E}_{i\text{-}j}(\Delta z) + \text{E}_{i\text{-}j}( \Delta z-g-L)$, where $0 \leq \Delta z < L+g$. For consistency with the standard definitions, we define $\text{D}$ in terms of $g$ such that $5\text{D} = L+g$. While this definition anticipates the value of $\text{D}$, it does not constrain it. 
Deferring the discussion of how $\text{D}$ is determined in simulations to the Appendix, we find that the values of $\text{D}$ that yield perfectly-staggered microfibrils fall within a narrow range, approximately $\text{D} \approx \SI[separate-uncertainty = true]{67(2)}{\nano\metre}$, see Fig. \ref{fig:boxWhisker_DBand}. Therefore, when we next discuss the minima of $\text{E}^{p}_{i\text{-}j}(\Delta z)$ over
$\Delta z$, we will use the corresponding values of $\text{D}$ as relevant length scales. 

\begin{figure*}[!t]
\centering
\makebox[\textwidth][c]{\includegraphics[scale=0.9]{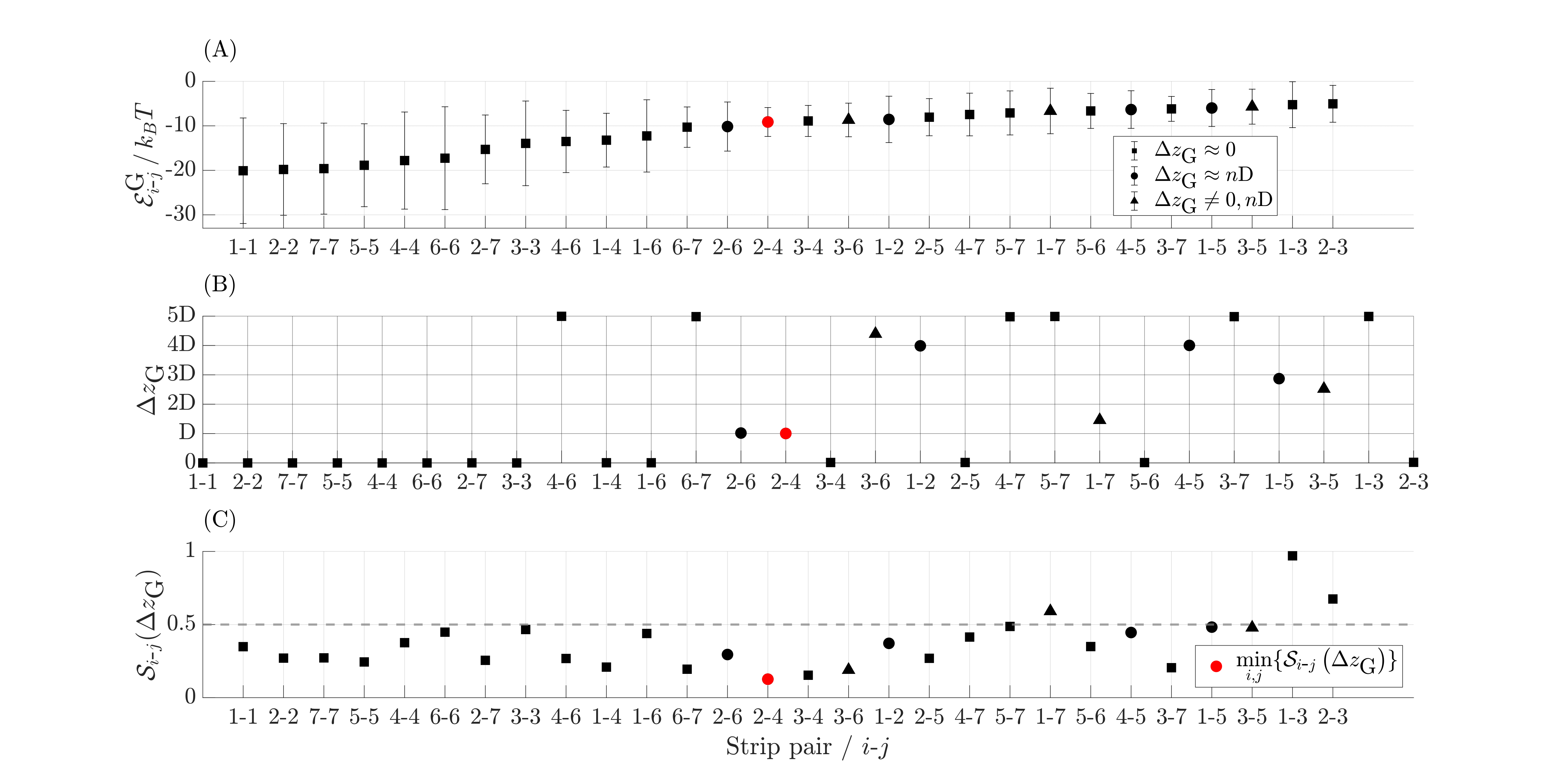}}%
\caption{Global minima of the axially periodic strip-strip interaction energies and their sensitivity to noise in contact potential values. The presented results are obtained numerically for $\alpha_{2}(\text{I})[\alpha_{1}(\text{I})]_{2}$ rat collagen. (A). Average values of the interaction energies at their global minima due to random noise in contact potential values. Error bars indicate the standard deviation in the noise-added energy values. (B). Locations of the global energy minima, $\Delta z_{
\textrm{G}}$. (C) Noise sensitivity of the global energy minima. The most pronounced minima belong to two classes: the minima at $\Delta z \approx 0$ and the minima at $\Delta z \approx \text{D}$. In general, the former are stronger but more sensitive to noise than the latter. See Materials and Methods for details of the procedures.}
\label{f:Ints_Example}
\end{figure*}

Fig. \ref{f:Ints_Example}B shows that the global minima of the strip-strip interaction potentials typically belong to two classes: the minima at $\Delta z \approx 0$ and the minima at $\Delta z \approx n\text{D}$. Motivated by the experimentally observed D-banded structures, previous studies have focused on local energy minima at positive multiples of $\text{D}$ overlooking the possibility of a global minimum at $\Delta z \approx 0$ \cite{hulmes1973analysis,trus1976molecular,piez1978sequence,hofmann1978role,puszkarska2022using}. Our results show that, unexpectedly, most, but not all, global minima fall into this class, see Figs. \ref{f:Ints_Example}A, B. This conclusion is surprising since the dominance of the minima at $\Delta z \approx 0$ would lead to an `in-register' arrangement of the molecules in a microfibril, precluding the formation of the D-staggered microfibril. This raises the question of what conditions warrant the formation of D-staggered microfibrils. 

To address this question, we note that the MJCP values used for calculating the interaction energies are subject to significant uncertainties due to experimental errors and high variability in biochemical environments. These uncertainties arise from neglecting the specifics of factors such as electrostatics, solvent effects, molecular crowding, and post-translational modifications (e.g., hydroxylation of Pro/Lys and glycosylation), as well as assuming sequence-independent interactions. Nevertheless, D-banded collagen fibrils do form under diverse conditions, including in {\em in vitro} environments (which lack biological regulatory factors). This suggests that the emergent structures must be highly robust toward environmental variability.

To account for it, we add random noise to the MJCP values and analyze the sensitivity of the intermolecular interactions and the emergent microfibrillar structures to this noise. We define the noise sensitivity of the pairwise strip-strip interactions as the variance of the noisy potentials $\prescript{}{}{{\mathcal{E}}}_{i\text{-}j}^{p}(\Delta z)$ normalised by their mean, i.e.
\begin{equation}
    \mathcal{S}_{i\text{-}j}(\Delta z) = \frac{\text{Var}\left\{\prescript{}{}{{\mathcal{E}}}_{i\text{-}j}^{p}(\Delta z)\right\}}{\left|\mu\left\{\prescript{}{}{{\mathcal{E}}}_{i\text{-}j}^{p}(\Delta z)\right\}\right|^{2}},
\end{equation}
where $\text{Var}\{\cdots\}$ and $\mu\{\cdots\}$ denote the variance and the mean, respectively. 
Fig. \ref{f:Ints_Example}C shows that, remarkably, the noise sensitivity turns out to be the smallest for the interaction potential minima at $\Delta z \approx \text{D}$. 
In contrast, the energy minima at $\Delta z\approx 0$ are more sensitive to the noise. 

We use the linear decomposition of $\mathcal{S}_{i\textrm{-}j}(\Delta z)$ into contributions from interacting pairs of residues (see Materials and Methods) to trace the high noise sensitivity of interactions at $\Delta z \approx 0$ to only two interacting pairs of highly abundant residues: Pro-Pro and Pro-Ala, see Fig. \ref{fig:NoiseContr}. The axial staggers and strip combinations that achieve the strongest inter-molecular interactions and minimise the number of interactions between abundant residues, end up being the least sensitive to the noise. 

In the biological context, the noise sensitivity implies that the majority of strip-strip interaction energies at $\Delta z\approx 0$ may be strongly affected by such factors as variations in pH and temperature or the post-translational hydroxylation of Pro residues \cite{bella2017fibrillar}. We hypothesize that this feature may form the basis of a sensitive biochemical control over the emergent structures. It requires a separate investigation in each biochemical context. For the present study, we simply assume that if the noise sensitivity of a minimum turns out to be higher than a chosen threshold $\mathcal{S}_{c}$, the minimum can be disregarded from the microfibril energy calculation. 

\subsection*{Emergence of D-periodic Microfibrils}
\begin{figure}[!ht]
\centering
\includegraphics[trim ={0cm 0cm 0cm 0cm}, clip, width = 0.7\linewidth]{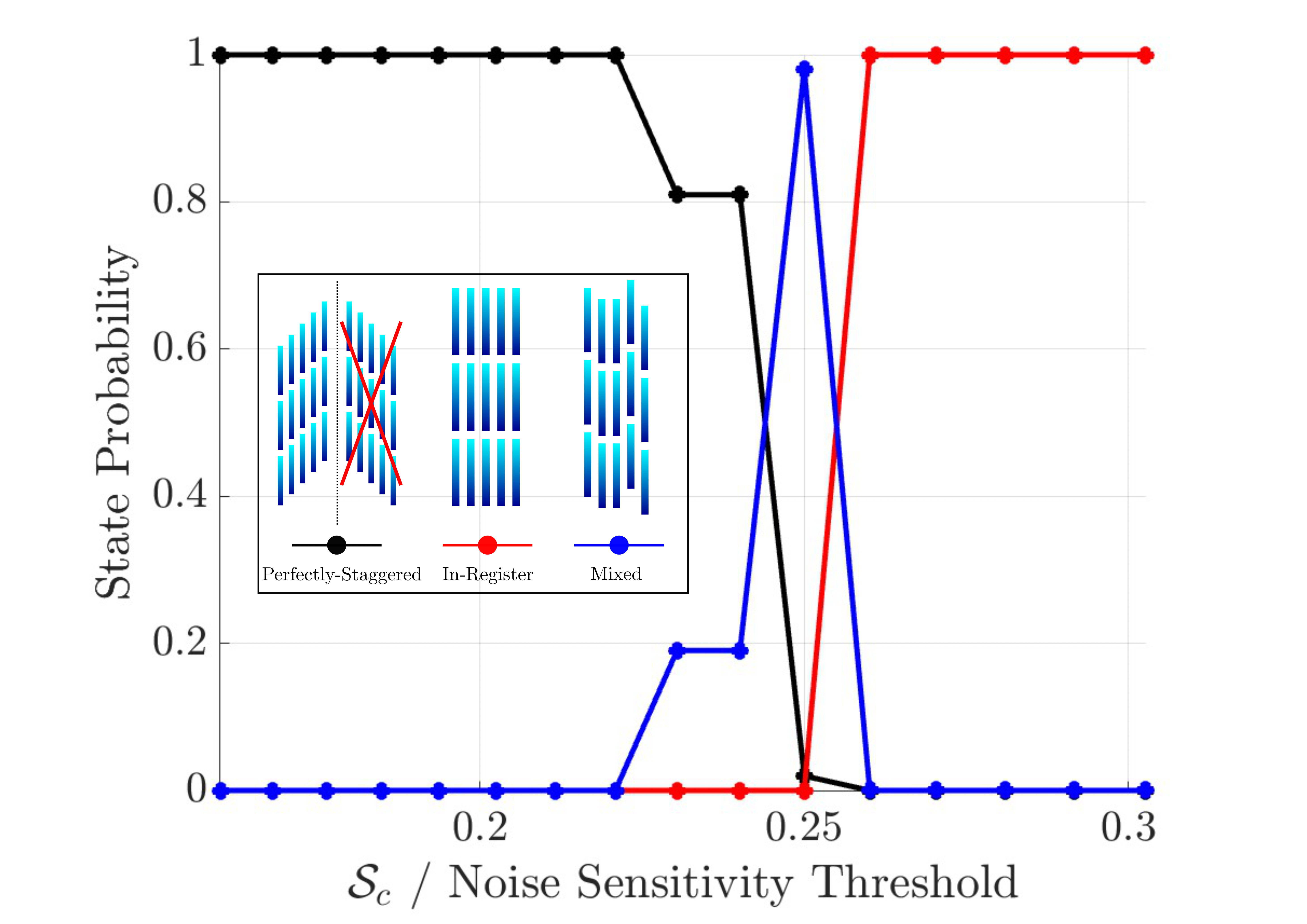}
\caption{Equilibrium probabilities of different microfibrillar states as a function of noise sensitivity threshold.}
\label{Microfibril_States}
\end{figure}

To determine whether the D-staggered microfibril emerges from the intermolecular interactions, elucidate the role of the strip-strip interactions, and explain why the microfibril is composed of $\text{N}=5$ molecules, we turn to numerical modeling. To keep the problem tractable, we assume an axially periodic microfibril comprised of aligned tropocollagen molecules placed at the vertices of a regular $\text{N}$-gon in the azimuthal plane. Individual molecules may rotate by the angles $\theta_{m}$ around their centerlines and be shifted along the microfibrillar axis by the distances $z_{m}$, see Fig. \ref{StructureFig}D. We assume that the only interacting molecules are the nearest neighbours that share a polygon edge. Any such pair of molecules is assumed to interact only via a single pair of strips at a time. The microfibril energy $\text{E}_{\textrm{M}}$ is then given by the sum of $\textrm{N}$ pairwise molecular interactions (see Materials and Methods). The equilibrium microfibrillar structure results from minimising the free energy of the system. Since the microfibril entropy is sub-extensive in microfibril length, it can be ignored in the present context and we can simply minimise $\text{E}_{\textrm{M}}$ \cite{puszkarska2022using}. 

For $\textrm{N}=5$, we identify three types of emergent microfibrillar configurations: (1) four perfectly-staggered ones, where each molecule is shifted relative to its right neighbour by the same value of $\Delta z \approx n\text{D}$ for $n = 1,2,3\,\ \text{or}\,\ 4$, (2) in-register configurations with $\Delta z \approx 0$ for all molecules, and (3) mixed configurations, see Fig. \ref{Microfibril_States}. When the sensitivity to the contact potential noise is disregarded, i.e. for high values of $\mathcal{S}_{c}$, simulations predict that the equilibrium microfibrils adopt the `in-register' configuration, consistent with the energetical dominance of the interaction minima at $\Delta z \approx 0$. As the acceptable noise sensitivity threshold is lowered, one of the perfectly-staggered configurations emerges at equilibrium, see Fig. \ref{Microfibril_States} and Fig. \ref{fig:perfectStagger_hist}. Perfectly-staggered microfibrils forming enantiomeric pairs ($\Delta z = \textrm{D}$ and $\Delta z = 4\textrm{D}$, $\Delta z = 2\textrm{D}$ and $\Delta z = 3\textrm{D}$) have differing energies in our model, so only one is selected at equilibrium. Without accounting for the chiral strip organisation of interacting residues, they would be energetically indistinguishable, see Materials and Methods.

\subsection*{Aggregate Size Specificity}
\begin{figure}[!b]
    \centering
    \includegraphics[ width = 0.8\linewidth,trim ={0 0 0cm 0cm}, clip]{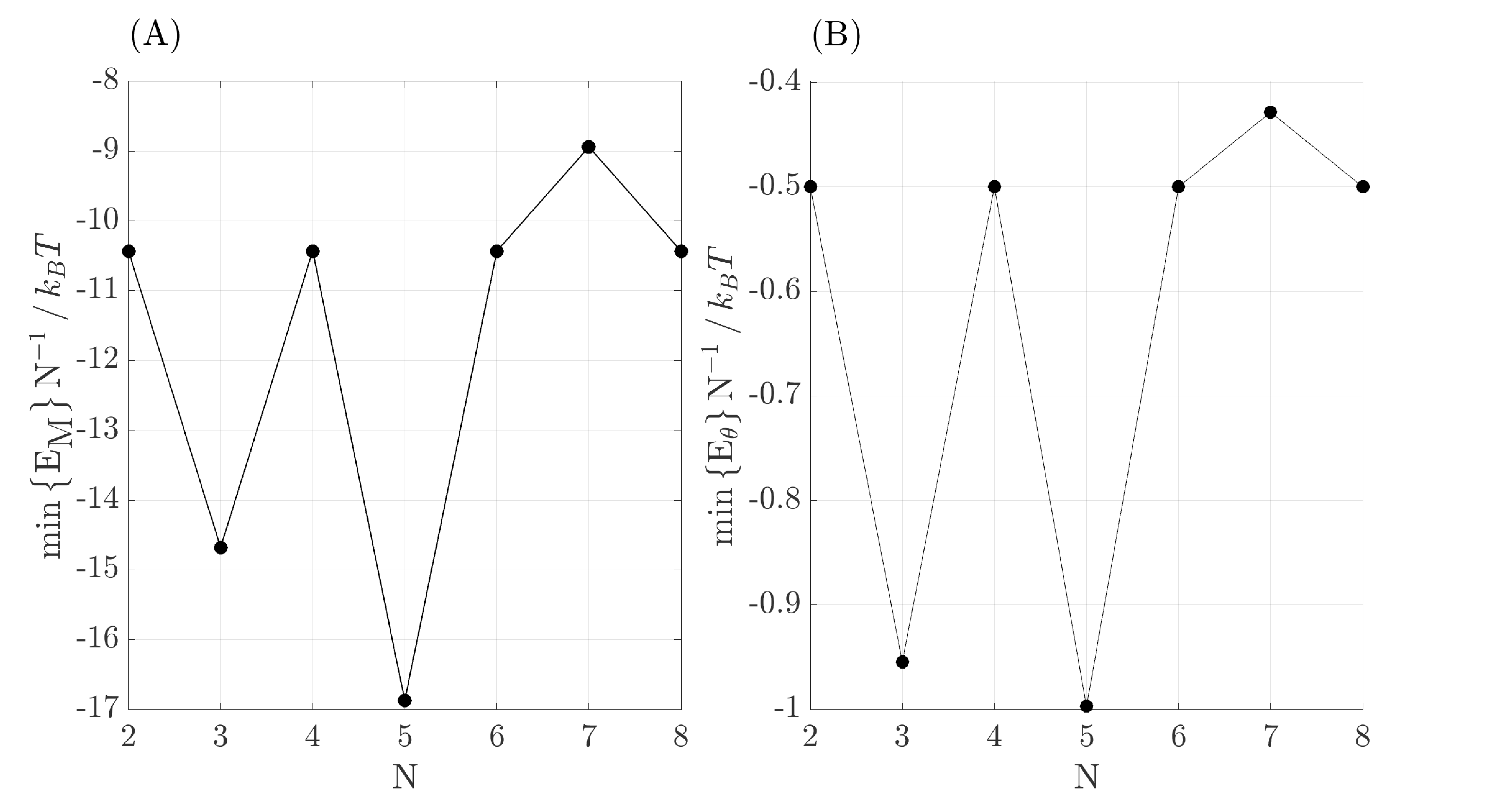}
    \caption{Global minimum of the microfibril energy per molecule as a function of cluster size N in $\alpha_{2}(\textrm{I})[\alpha_{1}(\text{I})]_{2}$ rat collagen. (A). Microfibril energy with empirically determined axial dependence. (B). Microfibril energy with no axial dependence. Details of the global optimisation procedure can be found in SI.}
    \label{EVsN}
\end{figure}

Next, we consider why a microfibril is comprised of five molecules. Fig. \ref{EVsN}A shows the global minimum of microfibrillar energy per molecule for varying $\text{N}$. Notably, $\text{N}=5$ gives the lowest energy, thus being selected at equilibrium. This fact has a simple geometrical explanation: for strong interactions, the helical strips of the neighbouring molecules must face each another. When molecules are positioned at the vertices of a regular $\text{N}$-gon, the interior polygon angle $v(\text{N})$ should approximate $m \alpha$, where $\alpha = 360^\circ/7$ is the angle between the strips and $m$ is an integer. It is easy to see that 
$v(5) = 108^\circ$ closely approximates $2 \alpha \approx 103^\circ$, see Table \ref{GeomArg}. For other values of $\text{N}$, some strips would always face away from their neighbours, reducing energetic gain. 

This argument relies on the spatial organisation of the residues in seven spirals, independent of their specific sequences. 
To substantiate this hypothesis, we perform simulations using intermolecular potentials that maintain azimuthal dependence due to the 7 helical strips but ignore axial dependence sensitive to sequence details. 
The results shown in Fig. \ref{EVsN}B indicate that the pentameric microfibril still has an energetic advantage over the trimeric aggregate, but this difference is reduced. Thus, although the spatial organisation of tropocollagen residues alone can make a pentamer the preferred microfibrillar configuration, specific residue interactions are essential for stabilising it. It is also conceivable that some specific residue sequences might preferentially select for a trimeric microfibril.   

\begin{table}[!h]
\centering
\begin{threeparttable}
\caption{Minimum difference between the internal angle of N-membered microfibrils and the azimuthal inter-strip spacings.}
\label{GeomArg}
\begin{tabular}{@{}lcc@{}}
\toprule
\hfill \textbf{Internal polygon angle} $v(\text{N})$ & \parbox{4.5cm}{\textbf{Minimum difference between} \\ $v(\text{N})$ \textbf{and $m\alpha$ / \text{deg}}} \\ 
\midrule
\hfil 60 &  8.6 \\
\hfil 90 &  12.9 \\ 
\hfil 108 & 5.1 \\ 
\hfil 120 & 17.1 \\ 
\hfil 128.6 & 25.7 \\
\hfil 135 & 19.3 \\
\bottomrule
\end{tabular}
\end{threeparttable}
\end{table}

\section*{Discussion}

The exquisite, axially periodic and helically entwined arrangement of collagen molecules in self-assembling fibrils and bundles of fibrils lies at the heart of collagen's versatility as a structural protein. This ordering emerges at the level of microfibrils -- essential, if experimentally elusive,  structures \cite{orgel2006microfibrillar}. In microfibrils,
five supercoiled molecules are staggered relative to their neighbours by a fixed distance D, and stacked to form an axially periodic structure. In this work, we investigate how amino acid sequences guide the formation of this structure. 

Focusing on collagen I, we found that the outward-facing tropocollagen residues are arranged in sevenfold helical strips. This arrangement emerges from the supercoiling of 
tropocollagen $\alpha$-chains and is reminiscent of the hydrophobic strip that emerges in coiled coils due to a regular spatial arrangement of heptad repeats \cite{mason2004coiled,neukirch2008chirality}. 
There are, however, important differences: there are seven, rather than just one, interaction strips, the residues forming the strips are, in general, not hydrophobic, and, most importantly, the seven-fold chiral arrangement emerges as a result of energetic selection favouring the spatial arrangement of residues described by the Pro-rich parametrisation of the tropocollagen triple helix. 

We predict that strip handedness and chirality are transmitted to the level of molecular collagen conformation through the torques that arise from pairwise intermolecular interactions. The resulting equilibrium helical angle $\phi^*$ is described by Eq. [\ref{e:phiStar}], which was first empirically obtained by Fraser and MacRae in the context of coiled coils \cite{fraser1973conformation}.

The stagger distance, $\text{D}\approx \SI{67}{\nano \metre}$, is encoded in local energy minima for the strip-strip interaction potentials, which occur at relative molecular stagger $\Delta z \approx n \textrm{D}, \quad n=1,\dots,4$ and at $\Delta z \approx 0$. While the minima at $\Delta z \approx 0$ are typically the strongest, they are sensitive to noise in the residue-residue interaction energies. This sensitivity is transmitted to the aggregate level. Upon introducing a noise sensitivity threshold that filters out noise-labile microfibrils, we find that the perfectly-staggered D-periodic microfibrils are the robust global minimisers of the microfibrillar free energy.  

\begin{table}[!t]
\centering
\begin{threeparttable}
\caption{Prediction of perfectly-staggered microfibrils in mammalian species for different collagen types.}
\label{Model_predict}
\begin{tabular}{@{}lcccc@{}}
\toprule
\hfil \textbf{Type} & \textbf{D-Banded Fibrils} &  \textbf{\textnumero \,\ Species} & \textbf{Predicted FS Microfibrils$  ^{\textbf{a}}$ (\%)} \\ 
\midrule
 \hfil $\alpha_{2}(\textrm{I})[\alpha_{1}(\text{I})]_{2}$ & \checkmark \cite{baselt1993subfibrillar} & \num{176} & \num{100}\\

\hfil $[\alpha_{1}(\textrm{I})]_{3}$ & \checkmark \cite{mcbride1997altered} & 186 & \num{25.27} \\ 

\hfil $[\alpha_{2}(\textrm{I})]_{3}$ & unknown$^{\textbf{b}}$  & 197 & \num{94.92} \\

\hfil $[\alpha_{1}(\text{II})]_{3}$ & \checkmark \cite{antipova2010situ}  & \num{191} & \num{100} \\ 

\hfil $[\alpha_{1}(\textrm{III})]_{3}$ & \checkmark \cite{asgari2017vitro,brodsky1980unusual}$^{\textbf{c}}$ & 185 & \num{3.78} \\ 

\hfil $[\alpha_{1}(\textrm{V})]_{3}$ & \text{\sffamily X} \cite{chanut2001control} & 167 & \num{86.83} \\ 

\hfil $\alpha_{1}(\text{V})\alpha_{2}(\text{V})\alpha_{1}(\text{V})$ & \checkmark \cite{mizuno2013fragility} & 151 & \num{78.81}\\

\hfil $\alpha_{1}(\text{V})\alpha_{3}(\text{V})\alpha_{2}(\text{V})$ & \checkmark \cite{mizuno2013fragility} & \num{124} & \num{98.39}\\  

\hfil $\alpha_{3}(\textrm{XI})\alpha_{2}(\textrm{XI})\alpha_{1}(\textrm{XI})$ & \checkmark \cite{hansen2003macromolecular}  & 148 & \num{99.32} \\ 

\hfil $[\alpha_{1}(\textrm{XXIV})]_{3}$ & unknown$^{\textbf{d}}$ & \num{163} & \num{87.12} \\

\hfil $[\alpha_{1}(\textrm{XXVII}]_{3}$ & \text{\sffamily X} \cite{plumb2007collagen}$^{\textbf{d}}$  & \num{163} & \num{60.36} \\
\bottomrule
\end{tabular}
\begin{tablenotes}
\small
\item $^{\textbf{a}}$ The model is said to predict a perfectly-staggered microfibril, if there exists some value of $\mathcal{S}_{c}$, below which the perfectly-staggered state probability is unity. 

\item $^{\textbf{b}}$ An $\alpha_{2}(\textrm{I})$ homotrimer is not observed \textit{in-vivo}. This homotrimer has been however observed in \textit{in-vitro} refolding experiments \cite{leikina2002type}. Propensity of $[\alpha_{2}(\textrm{I})]_{3}$ to undergo self-assembly into fibrils has not been investigated to our knowledge. 

\item $^{\textbf{c}}$ The D-banding length scale of reprecipitated type III collagen fibrils has been reported as \SI[separate-uncertainty = true]{66.7(0.2)}{\nano\metre} and \SI[separate-uncertainty = true]{25(10)}{\nano\metre}.

\item $^{\textbf{d}}$ Developmental collagens are characterised by presence of highly-conserved sequence interruptions. In this work, we do not account for their effect. 
\end{tablenotes}
\end{threeparttable}
\end{table} 

In the biophysical context, noise sensitivity points to a sensitive control that may be exerted on aggregates by local biochemical environments. For example, transitions between perfectly-staggered microfibrils and SLS aggregates (corresponding to the dominant minima at $\Delta z \approx 0$) can be induced by varying the interactions between charged residues \cite{xu2022segment}. Residue interactions may also be affected through post-translational enzymatic modification, such as hydroxylation of Pro and Lys residues. Post-translational hydroxylation is known to significantly affect the temperature and ionic conditions required for D-banded fibril formation \cite{perret2001unhydroxylated}. This observation aligns with our noise sensitivity analysis, which indicates that Pro-containing residue interactions, specifically Pro-Pro and Pro-Ala, have the largest effect on pairwise molecular energy.  

It has been understood at least since the work of Hodge and Petruska \cite{petruska1964subunit} and Smith \cite{smith1968molecular}, that to conform to the regular axial D-banded pattern, collagen aggregates must be composed of pentamers. However, to the best of our knowledge, the physical interactions that could warrant this have not been discussed. We find a strong energetic minimum at $\textrm{N}=5$ for an axially periodic $\textrm{N}$-membered microfibril. We show that for typical intermolecular interactions, the geometric condition that the strips of neighbour molecules face each other alone may select a pentameric microfibril. Yet, specific residue sequences are required for stabilising it and ensuring D-banding.

Our analysis relies on many assumptions and simplifications: we assume that the microfibril is a regular $\textrm{N}$-gon, that tropocollagen molecules are perfectly aligned, neglect the influence of collagen molecules external to the microfibril, and assume the validity of the contact potentials approach. Furthermore, we disregard the role of post-translational modifications, collagen telopeptide domains, and biological regulation among other factors. To validate our model, we have tested its predictions for all known mammalian sequences of fibril-forming collagens documented in the NCBI RefSeq database, see Table \ref{Model_predict}. 

High resolution \textit{in situ} studies of microfibrillar structures have only been performed for $\alpha_{2}(\text{I})[\alpha_{1}(\text{I})]_{2}$ collagen originating from rat tail tendon \cite{orgel2006microfibrillar}. Due to the limited availability of such detailed structural data for other collagen types, we use the available measurements of D-banding in different collagen types as a proxy for the emergence of the D-staggered microfibrils. Our analysis predicts that for all sequences of $\alpha_{2}(\textrm{I})[\alpha_{1}(\text{I})]_{2}$ collagen examined,  a perfectly-staggered microfibril is the most stable aggregate below some noise sensitivity threshold $\mathcal{S}_{c}$. This finding agrees with experimental observations of D-banding in collagens of this type. Similarly, our model indicates that over $99\%$ of the tested sequences for collagens  $[\alpha_{1}(\text{II})]_{3}$, $\alpha_{1}(\text{V})\alpha_{3}(\text{V})\alpha_{2}(\text{V})$ and $\alpha_{3}(\textrm{XI})\alpha_{2}(\textrm{XI})\alpha_{1}(\textrm{XI})$ favour the formation of perfectly-staggered microfibrils under the same conditions. 

However, our model's results for some homotrimeric collagens  $[\alpha_{1}(\textrm{I})]_{3}$, $[\alpha_{1}(\textrm{III})]_{3}$, $[\alpha_{1}(\textrm{V})]_{3}$ and $[\alpha_{1}(\textrm{XXVII})]_{3}$ seem to be at odds with the available experimental data. 
These discrepancies point to the limitations of the model assumptions used in this study and require future studies. For now, we note the possible sources of these discrepancies. First, in our model, pairwise molecular interactions between homotrimertic collagens of types I, III, and V are particularly strong, consistent with previous studies \cite{puszkarska2022using}.
This presents the tantalizing possibility that interactions of Pro-poor rather than Pro-rich strips might be selected despite the associated higher cost of elastic deformation. In addition to the selection of a different chiral symmetry, it is plausible that the molecular organisation or indeed the number of molecules comprising the microfibril may vary across different collagen types. In particular, specific residue sequences could in principle favour alternative microfibril configurations, such as a three-membered microfibril. In such cases, D-banding may first appear at the level of supramicrofibrillar structures, like pentameric aggregates composed of trimeric microfibrils. 

In our analysis of developmental collagens XXIV and XXVII, we did not account for sequence interruptions within their triple-helical domains. Studies using model peptides have demonstrated that deviations from the typical [Gly-X-Y] motif can cause localised unwinding or overwinding of the triple helix at the interruption site \cite{bella2010new}. These structural perturbations can significantly alter the amino acid composition of the helical strips following the interruption, thereby influencing the interaction potentials between these strips. Consequently, sequence interruptions can markedly affect the stability and assembly of collagen microfibrils \cite{hwang2010interruptions}. To our knowledge, the details of the structural impact of sequence interruptions present in collagens XXIV and XXVII remain to be elucidated \cite{bella2010new,boot2003novel,koch2003collagen}. This calls for future research into the effect of sequence interruptions in developmental collagens on spatial residue organisation and microfibril self-assembly. 

Finally, it should be noted that explicit measurements of D-banding have only been performed in a handful of commonly studied mammalian species. This raises the question of whether in certain species fibrillar collagens may aggregate into structures that lack D-banding. 

Understanding the differences in the self-assembly of heterotrimeric and homotrimeric collagens is crucial for uncovering the fundamental principles that underlie certain medical conditions. Deleterious mutations in the COL1A2 gene are known to lead to the production of homotrimeric type I collagen instead of the normal heterotrimeric form. Clinically, this mutation manifests itself as the Ehlers-Danlos syndrome, which is characterised by altered mechanical properties of collagen-contaning tissues, leading to joint hypermobility and cardiac valve abnormalities \cite{forlino2011osteogenImperfecta}. 

Our findings indicate that microfibrillar structural features are not uniform across all fibril-forming collagens. This is in good agreement with established knowledge in the field. 
For instance, corneal collagen fibrils exhibit a helical angle of approximately $\SI[]{15}{^{\circ}}$, significantly larger than that observed in bone or tendon tissues \cite{raspanti2018not}. Additionally, the characteristic periodic banding pattern can manifest at the fibrillar level with periodicities less than the typical D-spacing in type I and III collagens \cite{venturoni2003investigations, asgari2017vitro}. Recent advances in the synthesis of collagen-mimetic peptides also suggest the possibility of microfibril aggregate sizes differing from $\textrm{N} = 5$ \cite{Chen2019-ra}. 
We hypothesize that the nonuniformity of structural features among supramolecular collagen aggregates is a crucial characteristic that ensures collagen's structural versatility across diverse biological environments.
Studying these diverse self-assembly scenarios offers valuable opportunities for applying our theoretical methods to understand collagen structures. 

\section*{Conclusions}

This study identifies chiral intermolecular interactions, rooted in the spatial arrangement of outward-facing residues, as a fundamental mechanism driving the self-assembly of collagen into its characteristic D-periodic, supercoiled, pentameric microfibrils. By integrating residue-level sequence data with a physically motivated interaction model, we demonstrate that molecular chirality and microfibrillar architecture are intrinsically linked. The predicted right-handed supercoiling and staggered configuration are not only energetically favoured but also robust to biochemical noise across diverse mammalian collagen sequences. These insights bridge molecular sequence with mesoscale structure, offering a quantitative framework to understand fibrillar collagen assembly. Beyond elucidating a long-standing biophysical question, our approach provides guiding principles for the rational design of collagen mimetic materials, with potential applications in tissue engineering and synthetic extracellular matrices.

\section*{Materials and Methods}

\subsection*{Mechanism of Molecular Supercoiling and Energy-Driven Strip Selection}


We model the semi-flexible collagen molecule as an inextensible circular elastic rod, with residues organised on its surface in a family of helical strips with a pitch $h$. For a pair of molecules interacting via these helical strips, each molecule bends and twists into a (super)helical shape with the radius $R$ and a helical angle $\phi$ of the filament centreline.  This configuration aligns the residue strips of the interacting molecules to face toward other, incurring the elastic energy  
\begin{equation}
\label{el_energy}
\textrm{E}_{\textrm{el}} = \int_{0}^{L} \left( \frac{B}{2}\frac{\sin^{4}\phi}{R^{2}} + \frac{C}{2}\left(\frac{\sin2\phi}{2R}- \chi\frac{2\pi}{h}\right)^{2}\, \right) ds.
\end{equation}
Here $s$ is the arclength, $B$ and $C$ are the effective bending and twisting rigidities, respectively, $\chi=1$ for a right-handed strip while $\chi=-1$ for a left-handed strip \cite{neukirch2008chirality}. 
Taking $R$ to be fixed, for a sufficiently small ratio $\epsilon= 2\pi R / h \ll 1$ and a finite ratio $B/C$ (see Appendix for the derivation), leads to 
the equilibrium helical angle $\phi^{*}$ given by the asymptotic expression 
\begin{equation}
    \phi^{*} \approx \chi \epsilon = \chi \frac{2\pi R}{h}. 
    \label{e:phiStar}
\end{equation}

This is the classical Fraser and MacRae formula widely used to analyse the triple-helical structure of tropocollagen \cite{fraser1973conformation}. Neukirch {\em et al.} later extended it to include coiled-coil proteins under non-zero external forces and torques \cite{neukirch2008chirality}, albeit deriving it in a more restrictive limit where bending rigidity is much smaller than twisting rigidity $B/C \ll 1$, as opposed to a finite ratio of the two.  

Furthermore, we show that if, additionally,
\begin{equation}
    B/C \ll \epsilon^{-2},
    \label{e:cond2}
\end{equation}
the elastic equilibrium energy becomes dominated by the bending deformation component and is given by 
\begin{equation} \textrm{E}^{*}_{\textrm{el}} \approx 8 \pi^{4}
\frac{ B 
 R^{2} L }{h^{4}}
=8 \pi^{4} \frac{ \xi_{b}  R^{2} L }{h^{4}}k_{B}T
,   
\label{e:EbendAsympt}
\end{equation}
where $\xi_{b}$ is the bending persistence length. Condition [\ref{e:cond2}] is easily satisfied in practice.  
The expression [\ref{e:EbendAsympt}] shows that the energetic cost of the elastic deformation increases steeply as the helical pitch $h$ decreases. 

We estimate the molecular length as $L\sim \SI[]{300}{\nano\metre}$, the microfibril radius as $R \sim \SI[]{3}{\nano\metre}$, and the persistence length as $\xi_{b}\sim \SI[]{120}{\nano\metre}$ at neutral pH and physiological salt concentration \cite{rezaei2018environmentally}. For the Pro-rich strips with $h \sim \SI[]{200}{\nano\metre}$, the corresponding elastic energy cost is $\textrm{E}^{*}_{\textrm{el}} \sim$ \SI[]{0.16}{$k_{B}T$}. This value is significantly smaller than the characteristic interaction energy of two D-staggered collagen molecules, approximately \SI[]{10}{$k_{B}T$}. In contrast, for the Pro-poor strips with $h \sim \SI[]{75}{\nano\metre}$, the elastic energy cost is much higher at $\textrm{E}^{*}_{\textrm{el}} \sim$ \SI[]{8}{$k_{B}T$}, approximately $50$ times higher than that of the Pro-rich strips.

Thus, the interactions between outward-lying residues that cluster along spirals with the larger pitch $h$ are energetically strongly favoured.  Therefore, it is sufficient to account only for the interactions between the seven right-handed helical strips originating from the Pro-rich tropocollagen parametrisation when modelling the effective interaction potential collagen molecules. 
The equilibrium coiling angle for the corresponding helical pitch $h$ is estimated as $\phi^* \sim 5^\circ$, which aligns well with the experimental observation in bone and tendon \cite{baselt1993subfibrillar,raspanti2018not}. 

\subsection*{Axial Dependence of Pairwise Molecular Interactions}
In calculating pairwise molecular interactions, we will disregard the interactions involving the N/C-telopeptides, which whilst kinetically important, are not necessary for collagen self-assembly into D-banded fibrils \cite{kuznetsova1999does}. Denote a pair of interacting strips as $i\text{-}j$, wherein $i,j=1,2,\dots,7$. Let \{$\mathbf{e}_{\rho},\mathbf{e}_{\theta}, \mathbf{e}_{z}$\} be the set of cylindrical basis vectors in the triple helix coordinate system. Let $\mathbf{^{\mathrm{q}}x}_{j}$ to be the position vector of the residues along strip $j$ that are labeled in ascending order of axial coordinate by $\mathrm{q}\in\mathbb{Z}^{+}$. The staggered coordinates of $\mathbf{^{\mathrm{q}}x}_{j}$ are defined as
    \begin{equation}   \mathbf{^{\mathrm{q}}x}_{j}^{\prime}(\Delta z) = \mathbf{^{\mathrm{q}}x}_{j} + 2\pi h^{-1}\left(\Delta z + c_{j} - c_{i}\right)\mathbf{e}_{\theta} + \Delta z\mathbf{e}_{z},
\end{equation}
where $\Delta z$ is the axial stagger of strip $j$ relative to strip $i$ and $c_{l}$ are the constants that define the centerline equations of the strips $z = \frac{h\theta}{2\pi} + c_{l}$, for $l\in\{1,2,\dots,7\}$. The pairwise interaction energy for a staggered strip pair $i\text{-}j$ is then 
\begin{equation}
\label{energy_def}
    \text{E}_{i\text{-}j}(\Delta z) = \sum_{\mathrm{p},\mathrm{q}}\varepsilon_{g(\mathrm{p})g(\mathrm{q})}\left[\Theta\left(^{\mathrm{pq}}r_{ij}\right) - \Theta\left(^{\mathrm{pq}}r_{ij} - l_{c}\right)\right],
\end{equation} 
where $^{\mathrm{pq}}r_{ij} = |\mathbf{^{\textrm{p}}x}_{i}-\mathbf{^{\textrm{q}}x}_{j}^{\prime}|$,  $\Theta$ is the Heaviside step function, $l_{c}$ is the interaction length scale and $g:\mathbb{Z}^{+} \to \{1,2,\dots,20\}$ maps the sequential residue position along a strip onto its integer designation. 

The matrix $\varepsilon \in \mathbb{R}^{20\times20}$ represents the energies of the residue-residue interactions. We follow the method of Puszkarska \textit{et al.} and take the values of $\varepsilon$ to be the empirically determined Miyazawa-Jernigan contact potentials, namely the entries MIYS850103, MIYS960102, MIYS990107 in the AAIndex database \cite{kawashima2000aaindex}. We take $l_{c} = \SI[]{0.75}{\nano\metre}$, which is typically assumed to be the representative length scale at which a pair of residues is in contact \cite{puszkarska2022using}.

Interactions between axially periodic arrays of parallel collagen molecules separated by gaps of length $g$ are described by the 28 $T$-periodic potentials, where $T=L+g$: 
\begin{equation}
\label{eqn:Ep_defn}
\text{E}^{p}_{i\text{-}j}(\Delta z) = \text{E}_{i\text{-}j}(\Delta z) + \text{E}_{j\text{-}i}(T-\Delta z),
\end{equation}
where $0 \leq \Delta z < T$. When $i=j$, the sequences of the opposing strips are identical and 
\begin{equation}
\text{E}^p_{i\text{-}i}(\Delta z) =  \text{E}^p_{i\text{-}i}(T-\Delta z),
\end{equation}
i.e. the functions $\text{E}^p_{i\text{-}i}$ possess a reflection symmetry with respect to $\Delta z = T/2$. Since previous studies \cite{puszkarska2022using} did not differentiate between residue strips, their interaction potentials inherently exhibit this property. In particular, this means that such physical interactions do not distinguish between the enantiomeric pairs corresponding to $\Delta z$ and $T-\Delta z$. This property might lead to a degenerate ground state, precluding formation of a well-defined axially periodic (D-banding) structure. In particular, perfectly-staggered right-handed and left-handed microfibrils corresponding to the symmetric values of $\Delta z$ could not be differentiated.   
This symmetry is broken for interactions of different  strips,
\begin{equation}
\text{E}^p_{i\text{-}j}(\Delta z) \ne  \text{E}^p_{i\text{-}j}(T-\Delta z), \quad i \ne j,
\end{equation}
lifting the degeneracy.

\subsection*{Noise Sensitivity of Pairwise Interactions}
To account for uncertainty in the elements of $\varepsilon$, consider a noise-added residue interaction matrix with elements $\varepsilon^{*}_{lm} = \varepsilon_{lm} + u_{lm}$. We choose $u_{lm}\sim U(a,b)$, where $U(a,b)$ is the continuous uniform distribution on the interval $(a,b)$. The noise-added pairwise interaction energy $\mathcal{E}^{p}_{i\textrm{-}j}$ is then calculated according to Eq. [\ref{energy_def}] and Eq. [\ref{eqn:Ep_defn}] using the matrix $\varepsilon^{*}$. 

The noise sensitivity parameter can be analytically evaluated for $\mathcal{E}^{p}_{i\textrm{-}j}$ using the following expression 
\begin{equation}  
\label{analytic_Sij}
\mathcal{S}_{i\textrm{-}j}(\Delta z) = \frac{12^{-1}\left|a-b\right|^{2} \sum\limits_{\mathclap{1\leq l \leq m \leq 20}} N_{lm}^{2}}{\left|\textrm{E}_{i\text{-}j}(\Delta z) + \frac{1}{2}\left(a+b\right)\sum\limits_{\mathclap{1\leq l \leq m \leq 20}} N_{lm} \right|^{2}},
\end{equation}
where $N_{lm}$ is the number of interacting residues with integer designations $l$ and $m$ at a given $\Delta z$. Importantly, Eq. [\ref{analytic_Sij}] is a linear combination of the contributions due to pairs of interacting residues proportional to $N_{lm}^2$. 

In addition to the analytical expression in Eq. [\ref{analytic_Sij}], the noise sensitivity parameter can be computed numerically. The results presented in Fig. \ref{f:Ints_Example} were performed numerically by constructing $K = 50$ noise-added interaction energy curves, with noise sampled from $U(-0.1 k_B T,0.1 k_B T)$. The value of the noise amplitude $\left|a \right| (= \left|b \right|)$ is unknown, as such for convenience we chose it to be $\approx 10 \%$ of the maximum value of the matrix elements in $\left|\varepsilon\right|$. Importantly, the relative noise sensitivity of the energies and, hence, all of our physical conclusions are independent of the chosen value. 

\subsection*{Model of a Microfibril} We parametrise the azimuthal component of pairwise molecular energy by 
\begin{equation}
    \Phi(\theta_{m}) = \left[1 + \exp\left(a\left|\sin\left(\frac{\pi(\theta_{m}-\theta_{0})}{\theta_{\text{\text{d}}}}\right)\right|-b\right)\right]^{-1},
\end{equation}
where the parameters $\theta_{0},\theta_{\text{d}},a,b$ are chosen to produce 7 equally-spaced maxima for $\theta_{m}\in [0,2\pi)$ with the same width as the Pro-rich strips (further details can be found in SI). The pairwise molecular energy can be written as 
\begin{align}
    P_{m} = \Phi(\theta_{m})\Phi(\theta_{m+1}-v)\text{E}^{p}_{n(\theta_{m})\text{-}n(\theta_{m+1}-v)}(\Delta z_{m}),
\end{align}
where $n(\theta_{m}) = \text{nint}\left[\left(\theta_{m}-\theta_{0}\right)\theta_{d}^{-1}\right]\bmod{7} + 1$, $\text{nint}[\cdots]$ rounds its argument to the nearest integer and $v$ is the internal angle of the N-gon. The energy of the whole microfibril is then simply 
\begin{equation}
    \text{E}_{\text{M}} =  \sum_{m=1}^{\text{N}-1}P_{m} + \Phi(\theta_{\text{N}})\Phi(\theta_{1}-v)\text{E}^{p}_{n(\theta_{\text{N}})\text{-}n(\theta_{1}-v)}(\Delta z_{\text{N}}).
\end{equation}
Cyclical connectivity of the N-gon constrains $\Delta z_{\text{N}} \equiv z_{1} - z_{\text{N}} = -\sum_{m=1}^{\text{N-1}}\Delta z_{m}$, where $\Delta{z}_{m} = z_{m+1} - z_{m}$ is the relative axial translation between two molecules.

\section*{Author Contributions}
A.Z., R.K. and D.O.P. conceptualised the project. A.Z. and D.O.P. developed theoretical models. A.Z. carried out empirical and numerical analysis. A.Z., R.K. and D.O.P. wrote the paper.

\section*{Appendix}
\subsection*{Detailed Parametrisation of the Azimuthal Energy Component}
Parameters $a$ and $b$ that are used in the definition of the azimuthal energy component $\Phi$ are parametrised as follows: 
\begin{align}
    b &= \frac{\log(q-1)f(\theta_{\text{f}}) - \log(p-1)f(\theta_{\text{w}})}{f(\theta_{\text{w}})-f(\theta_{\text{f}})}, &
    a &= \frac{\log(q-1)+b}{f(\theta_{\text{w}})}, & f(t) &= \sin\left(\frac{\pi t}{2\theta_{\text{d}}}\right).
\end{align}
Parameters $(\theta_{\text{w}},\theta_{\text{f}},p,q)$ are defined via
\begin{equation}
\Phi(\theta_{\text{max}}\pm\theta_{\text{f}}) = p^{-1}, \,\ \Phi(\theta_{\text{max}}\pm\theta_{\text{w}}) = q^{-1},
\end{equation}
where $\theta_{\text{max}}$ maximises $\Phi(\theta_{m})$ for $\theta_{m}\in[0,2\pi)$. In all calculations we set 
\begin{equation}(\theta_{0},\theta_{\text{d}},\theta_{\text{w}},\theta_{\text{f}},p,q) = \left(0.5,\frac{2\pi}{7},\frac{\pi}{9},0.06,1.0004,100\right).
\end{equation}

\subsection*{Global Optimisation of Microfibrillar Energy \& Calculation of Equilibrium Statistics} 

In this section we outline the algorithm for global optimisation of the microfibril energy and subsequent calculation of equilibrium microfibril statistics. 

\subsubsection*{Selection of the D-banding Lengthscale} The first step in calculating the possible values of the microfibril energy is deciding on the value of the D-banding lengthscale. This then allows for construction of axially-periodic pairwise potentials $\textrm{E}^{p}_{i\textrm{-}j}$, which determine the value of the microfibril energy - see equation [15] of the main text. A priori we do not know the exact value of the parameter D. We start by constraining $\textrm{D}\in[620,700]\textrm{\AA}$, based on the experimental measurements of D-banding \cite{bella2017fibrillar,chen2017relationship}. Next, for each amino acid sequence, we construct a set of candidate values for the D-banding lengthscale, based on the axial stagger of the interaction energy minima of non-periodic pairwise potentials $\textrm{E}_{i\textrm{-}j}$. The set of candidate values for D is defined as 
\begin{equation}
    S_{\textrm{D}} = \left\{\Delta\widetilde{z} \in[620,720]\textrm{\AA} \,\bigg\vert\, \Delta\widetilde{z} = \underset{\Delta z}{\arg\min}\{\textrm{E}_{i\textrm{-}j}(\Delta z)\},\,  \mathcal{S}_{i\textrm{-}j}(\Delta\widetilde{z}) < \mathcal{S}_{\textrm{thr}} \right\},
\end{equation}
where $\mathcal{S}_{\textrm{thr}} = 0.49$ is the threshold value of noise sensitivity, below which the minimum is considered a candidate value. $\mathcal{S}_{\textrm{thr}}$ serves as means of roughly filtering out candidate values of D that are unlikely to give rise to interactions with low noise sensitivity. For practical calculations, we restrict the number of elements in $S_{\textrm{D}}$ by further requiring that $\Delta\widetilde{z}$ only correspond to global, secondary or tertiary minima of the pairwise potentials $\textrm{E}_{i\textrm{-}j}$.

We can now construct a numerical grid of candidate $\textrm{D}$ values to be used for further calculations. The grid points are sampled from 
\begin{equation}
    I_{\textrm{D}} = \bigcup\limits_{\Delta\widetilde{z}\in S_{\textrm{D}}}\left[{\Delta \widetilde{z}} - \delta z,\Delta\widetilde{z} + \delta z\right],
\end{equation}
where we pick $\delta z = \textrm{\SI[]{3}{\angstrom}}$. We sample candidate values of D by first discretising each closed interval comprising $I_{\textrm{D}}$ into a uniformly-spaced grid with spacing of \SI[]{0.5}{\AA}. If we have an overlap between intervals, we pick the grid points for the discretisation that are associated with the least noise sensitive $\Delta\widetilde{z}$. We now construct axially-periodic pairwise potentials $\textrm{E}^{p}_{i\textrm{-}j}$ using  a candidate D value that is generated from $\Delta\widetilde{z}$ with the lowest noise sensitivity. 

\subsubsection*{Construction of Near-Equilibrium States}
The next step is constructing an approximation to the spectrum of the microfibril. Studying the predictions of our model at thermal equilibrium necessitates performing global optimisation of the microfibril energy $\text{E}_{\text{M}}$. An N-membered collagen microfibril has $2\text{N}-1$ degrees of freedom in our model. To aid us in finding the global minimum of $\text{E}_{\text{M}}$, we construct near-equilibrium states (NEqS) which will give the largest energy contributions to the spectrum. NEqS are members of the set $\text{S}_{\text{eq}} = 
\{(\boldsymbol{\theta}^{\text{eq}},\boldsymbol{\Delta z}^{\text{eq}})\}$, in which the pair of state vectors $(\boldsymbol{\theta}^{\text{eq}},\boldsymbol{\Delta z}^{\text{eq}})$ specifies the microscopic state of a microfibril. The components of azimuthal state vector maximise the strip overlap in a given $\text{N}$-gon:  
\begin{equation}
\label{azVec}
    \theta_{l}^{\text{eq}} \in \underset{\theta\in[0,2\pi)}{\arg\max} \{  \Phi(\theta)\Phi(\theta-v)\},\,\ l = 1,\dots,\text{N}.
\end{equation}
The axial state vector contains $\text{N}-1$ components which correspond to the staggers that minimise the axial energy component for a given pair of strips in a microfibril. The total number of NEqS is $7^{\text{N}}\text{M}^{\text{N}-1}$, where M is the number of minimisers for each interaction curve $\text{E}^{p}_{i\text{-}j}$. To keep the problem tractable, we choose $\text{M} = 3$. 

In a given microfibril the axial energy components of $P_{m}$ in general will not be the same. To account for this, we relax the azimuthal degrees of freedom using the sequential quadratic programming algorithm over the domain $[\theta_{1}^{\text{eq}}-\delta\theta,\theta_{1}^{\text{eq}}+\delta\theta]\times\dots\times[\theta_{\text{N}}^{\text{eq}}-\delta\theta,\theta_{\text{N}}^{\text{eq}}+\delta\theta]$ with $\delta\theta = 0.15$.

\subsubsection*{Calculation of Equilibrium Probabilities with Noise Sensitivity} Finally, we calculate the equilibrium statistics of the collagen microfibril. We group the microscopic microfibril states into 3 categories based on the components of $\boldsymbol{\Delta z}^{\text{eq}}$. The definitions of the microfibril categories are shown in Table \ref{MicroCategs} and Fig. \ref{Microfibril_States} of the main text. Let $s_{k}$ denote the $k^{\text{th}}$ microfibril in one of these three categories, which we will denote by $s\in\{A,B,C\}$. The equilibrium probability in the canonical ensemble formalism for perfectly-staggered microfibrils (category $A$) is then 

\begin{table}[t]
\centering
\begin{threeparttable}
\caption{Definitions of microfibril categories, based on the pattern of axial staggers.}
\label{MicroCategs}
\centering
\begin{tabular}{@{}lccc@{}}
\toprule
\hfill \textbf{Microfibril Category} & \textbf{Category Index} & $^{\textbf{a}}$\textbf{Condition on Axial Staggers} \\
\midrule
\hfill Perfectly-staggered & $A$ & $\Delta z_{m}^{\textrm{eq}} = n\textrm{D}, \,\ \text{for  all} \,\ m = 1,\dots,\textrm{N}-1$ \\
\midrule 
\hfill In-register & $B$ & 
$\Delta z_{m}^{\textrm{eq}} \in [-\Delta_{0},\Delta_{0}] \,\ \text{for  all} \,\ m = 1,\dots,\textrm{N}-1$ \\
\midrule
\hfill Mixed & $C$ & Any other $\Delta z_{m}^{\textrm{eq}}$ that are not perfectly-staggered or in-register \\ 
\bottomrule
\end{tabular}
\begin{tablenotes}
\small 
\item $^{\textbf{a}}$ We set the parameter $\Delta_{0} = \SI[]{5}{\nano\metre}$\ and $n = 1,2,3$ 
or 4
\end{tablenotes}
\end{threeparttable}
\end{table}

\begin{equation}
    \mathcal{P}_{A} = \frac{\sum\limits_{k} \exp(-\beta\prescript{A_{k}}{}{\text{E}_{\text{M}}^{\text{eq}}})}{\sum\limits_{s}\sum\limits_{k} \exp(-\beta\prescript{s_{k}}{}{\text{E}_{\text{M}}^{\text{eq}}})},
\end{equation}
where $\prescript{A_{k}}{}{\text{E}_{\text{M}}^{\text{eq}}}$ is the microfibrillar energy of the $k^{\textrm{th}}$ perfectly-staggered microfibril and $\beta^{-1} = k_{B}T$. Analogous formulae define the equilibrium probabilities of mixed and in-register states. 

For a given noise sensitivity threshold $\mathcal{S}_{c}$, we include a NEqS in calculation of $\mathcal{P}_{s}$ if for a given $\boldsymbol{\theta}^{\text{eq}}$, the components of the stagger vector satisfy
\begin{equation}
\label{noise_thresh_cond}
    \mathcal{S}_{i\text{-}j}(\Delta z_{l}^{\text{eq}}) < \mathcal{S}_{c} \,\ \text{for all} \,\ l = 1,\dots,\text{N},
\end{equation}
where $\Delta z_{\text{N}} = \left(-\sum\limits_{m=1}^{\text{N}-1}\Delta z_{m}^{\text{eq}}\right)\bmod 5\text{D}$. We note that Eq. [\ref{noise_thresh_cond}] must hold for all strip pairs $i$-$j$ which interact in a microfibril specified by the azimuthal state vector $\boldsymbol{\theta}^{\text{eq}}$. 

If we find that there exists a value of $\mathcal{S}_{c}$, such that $\mathcal{P}_{A}\to 1$, we say that our model predicts perfectly-staggered microfibrils at thermal equilibrium. If such a value of $\mathcal{S}_{c}$ does not exist, we repeat our calculations with a different candidate value for the D-banding lengthscale. If all such candidate values are exhausted, we conclude that perfectly-staggered microfibrils are not expected at equilibrium within our modelling framework. 

\subsection*{Derivation of the Asymptotic Expression for Equilibrium Helical Angle $\phi^{*}$}
Let us assume that the supercoiling radius $R$ has a fixed value and the helical angle $\phi$ is independent of the arcelength $s$ in Eq. [2] of the main text. Then, the elastic energy of deformation is minimised for an equilibrium helical angle $\phi = \phi^{*}$ which satisfies
\begin{equation}
\label{eqn:eqbmCond}
2\gamma\sin^{3}\phi^{*}\cos\phi^{*} + (\sin\phi^{*}\cos\phi^{*} - \chi\epsilon)\cos2\phi^{*} = 0,
\end{equation}
where we have defined $\gamma = B/C$ and $\epsilon = 2\pi R/h$. 

Our goal is to construct an asymptotic series for the equilibrium helical angle $\phi^{*}$ as a function of $\epsilon\ll 1$ and finite $\gamma$. To that end, we note that we can write $\epsilon$ as a function of $\phi^{*}$ in Eq. [\ref{eqn:eqbmCond}], obtaining
\begin{equation}
    \overline{\epsilon} = (1-\gamma)\sin\overline{\phi} + \gamma\tan\overline{\phi} \equiv f\left(\overline{\phi}\right),
\end{equation}
where for convenience we have defined $\overline{\phi} = 2\phi^{*}$ and $\overline{\epsilon} = 2\chi\epsilon$. The desired asymptotic expression for $\phi^{*}$ is therefore equivalent to finding the series expansion of the inverse function $g\equiv f^{-1}$. Noting that $f$ is analytic at $\overline{\phi} = 0$ and that $f^{\prime}(0) = 1$, we can apply the Lagrange inversion formula \cite{de2014asymptotic} to obtain 
\begin{equation}
    \overline{\phi} \equiv g\left(\overline{\epsilon}\right) = \sum_{n=1}^{\infty}g_{n}\overline{\epsilon}^{n}, 
\end{equation}
where the expansion coefficients are given by 
\begin{equation}
\label{eqn:hns}
g_n = \frac{1}{n!}\lim_{\overline{\phi}\to 0}\frac{\mathrm{d}^{n-1}}{\mathrm{d}\overline{\phi}^{n-1}}\left(\frac{1}{h\left(\overline{\phi}\right)}\right)^{n},\,\ \text{where}\,\  h\left(\overline{\phi}\right) = \frac{f\left(\overline{\phi}\right)}{\overline{\phi}}.
\end{equation}
We note that for $\overline{\phi}\ll 1$ we can expand $h\left(\overline{\phi}\right) = 1 + \frac{3\gamma-1}{6}\overline{\phi}^{2} + O\left(\overline{\phi}^{4}\right)$. Using Eq. [\ref{eqn:hns}], the equilibrium helical angle is then asymptotically found to be
\begin{equation}
\label{eqn:phistar}
    \phi^{*} = \chi\epsilon + \frac{2(1-3\gamma)}{3}(\chi\epsilon)^{3} + O\left(\epsilon^{5}\right).
\end{equation}
With the aid of the asymptotic expression in Eq. [\ref{eqn:phistar}], we can estimate the equilibrium bend and twist energy contributions per unit length as 
\begin{align}
    \textrm{E}_{\textrm{bend}} &= \frac{B}{2}\frac{\sin^{4}\phi^{\star}}{R^{2}} \sim \frac{B\epsilon^{4}}{2R^{2}}, & \textrm{E}_{\textrm{twist}} &= \frac{C}{2R^{2}}\left(\frac{\sin2\phi}{2}- \epsilon\right)^{2}\ \sim \frac{C\gamma^{2}\epsilon^{6}}{2R^{2}}.
\end{align}
We therefore conclude that in the limit $\epsilon^{-2}\gg \gamma$, the bend contribution to the total equilibrium elastic deformation energy is dominant over the twist contribution. 


\setcounter{figure}{0}
\renewcommand{\thefigure}{S\arabic{figure}}

\newpage
\begin{figure}
\centering
\includegraphics[width=1\textwidth, trim ={0cm 0cm 0cm 0cm}, clip, width = \linewidth]{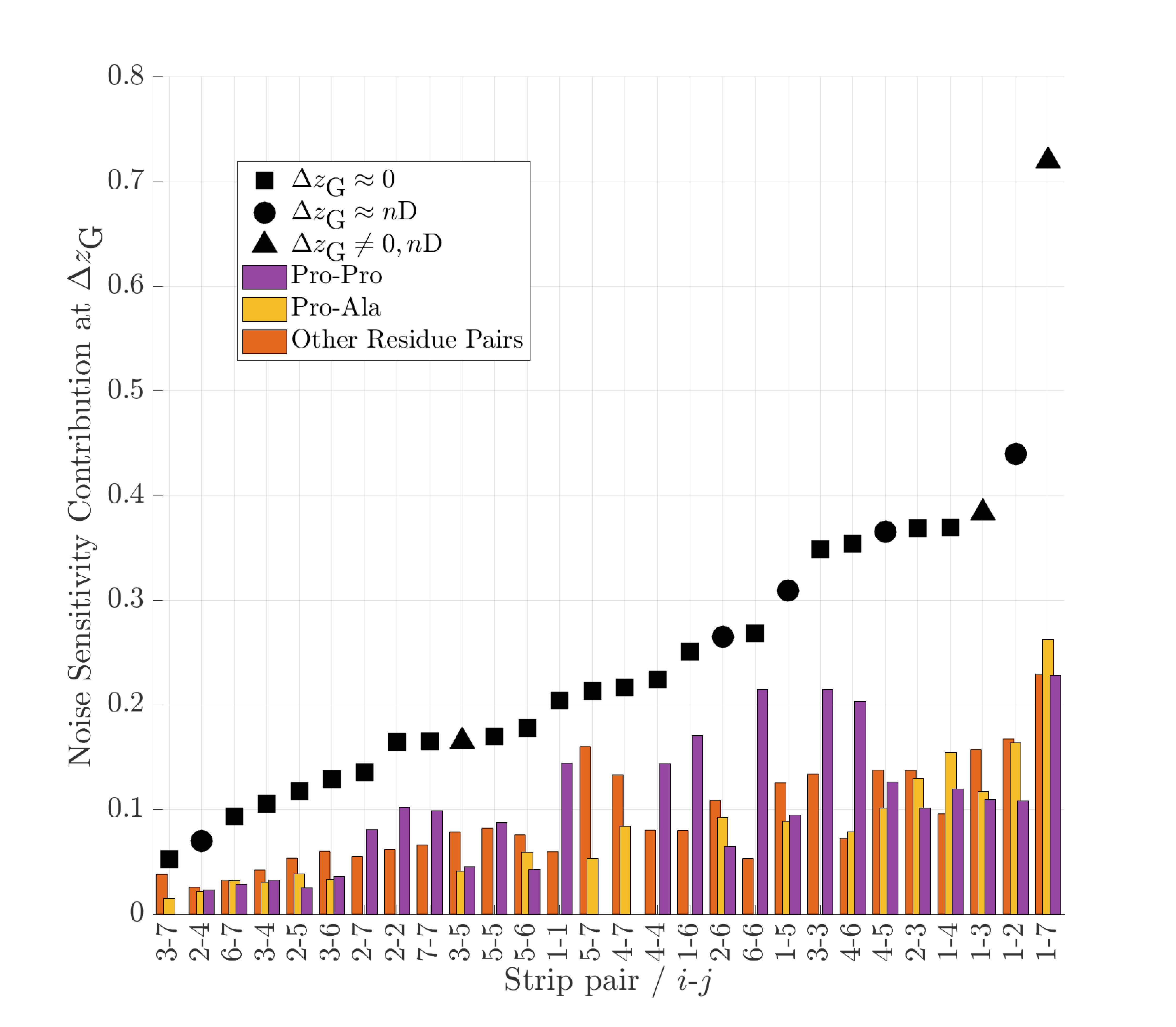}
\caption{The largest contributions to the noise sensitivity of the global minima of the strip-strip energy $\textrm{E}_{i\textrm{-}j}^{p}(\Delta z_{\text{G}})$ from the pairs of interacting residues. The black markers show the total noise sensitivity of each minimum.  All residue pairs which contribute less than $20\%$ to the total noise sensitivity, are categorised as `other residue pairs' (see Materials and Methods for details). Typically, the highest contributions to the noise sensitivity come from two pairs of interacting residues, Pro-Pro and Pro-Ala.}
\label{fig:NoiseContr}
\end{figure}

\newpage
\begin{figure}
\centering
\includegraphics[width=1\textwidth, trim ={0cm 0cm 0cm 0cm}, clip, width = \linewidth]{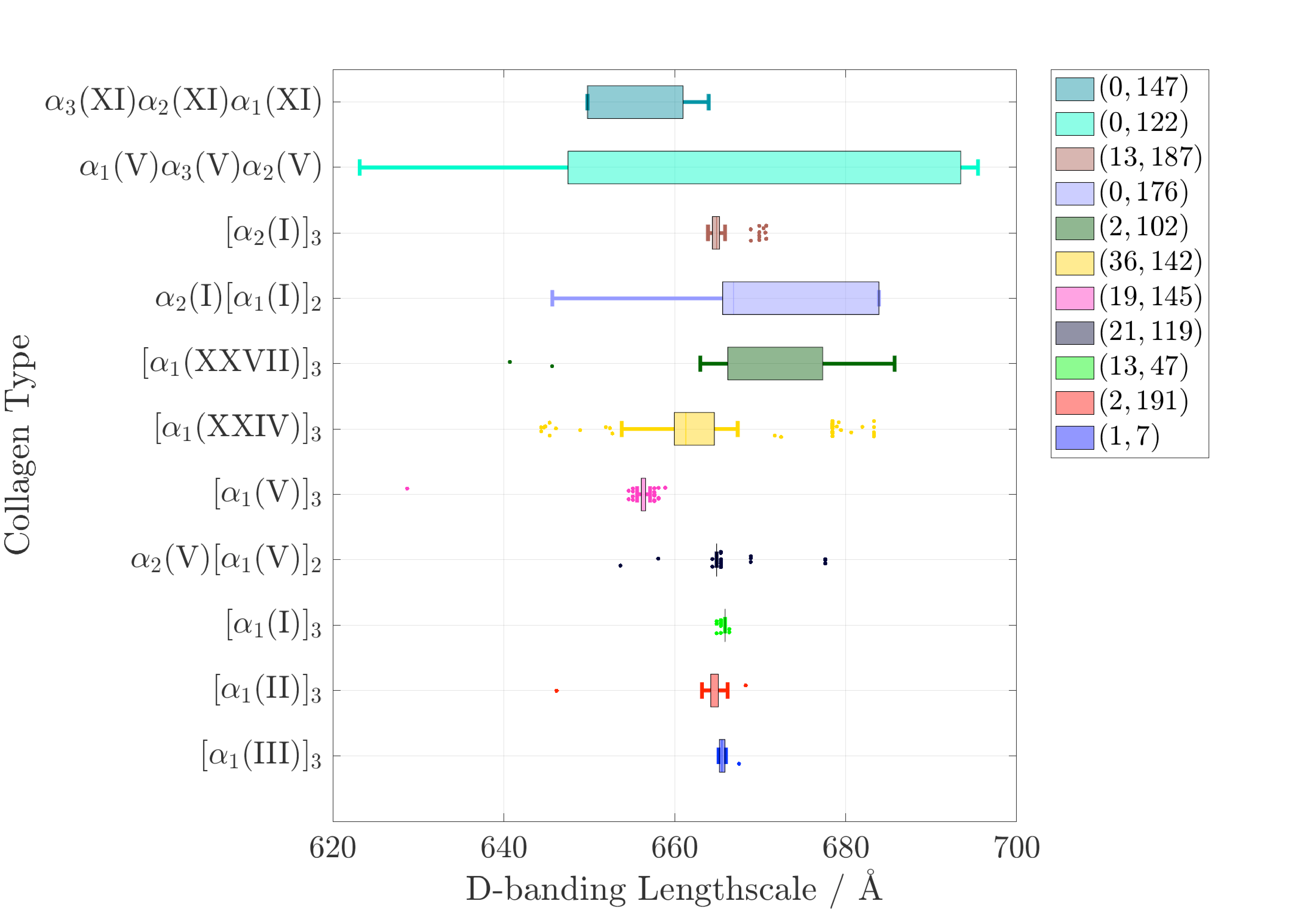}
\caption{Box plot of the D-banding lengthscales across different mammalian species that give rise to stable perfectly-staggered microfibrils. The whiskers are drawn up to the largest/smallest data point that is within  $1.5\cdot \textrm{IQR}$ (inter-quartile range) of the upper/lower quartile, indicated by the top/bottom edges of each box respectively. D-banding lengthscales that are a distance more than $1.5\cdot \textrm{IQR}$ from the top/bottom of a box are labelled as outliers and plotted as individual points. To each box, we associate an ordered pair $(N_{\text{out}},N_{\text{tot}})$, where $N_{\text{out}}$ denotes the number of outliers and $N_{\text{tot}}$ denotes the total number of species for which stable perfectly-staggered microfibrils were found.}
\label{fig:boxWhisker_DBand}
\end{figure}

\newpage
\begin{figure}
\centering
\includegraphics[width=1\textwidth, trim ={0cm 0cm 0cm 0cm}, clip, width = \linewidth]{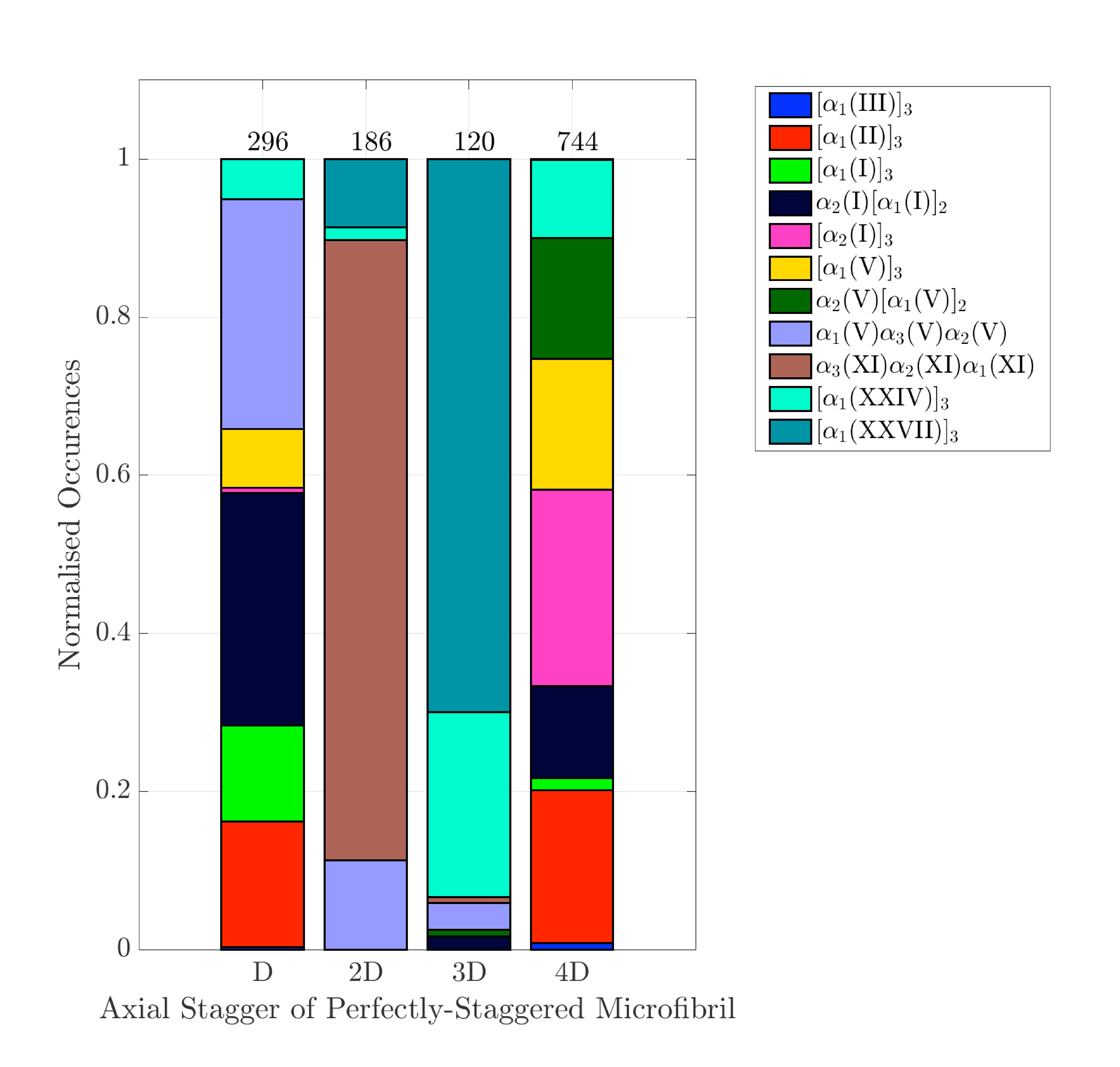}
\caption{Histogram of the axial stagger value in stable perfectly-staggered microfibrils across different collagen types in mammalian species. The number of occurrences is normalised by the total number of stable perfectly-staggered microfibrils with a given axial stagger, which is shown at the top of each bar. A microfibril is deemed perfectly-staggered, provided that each axial stagger is within $5\%$ of the same integer multiple of the D-banding lengthscale (values used are those shown in Figure \ref{fig:boxWhisker_DBand}). }
\label{fig:perfectStagger_hist}
\end{figure}

\newpage
\begin{figure}
\centering
\includegraphics[width=1\textwidth, trim ={0cm 0cm 0cm 0cm}, clip, width = \linewidth]{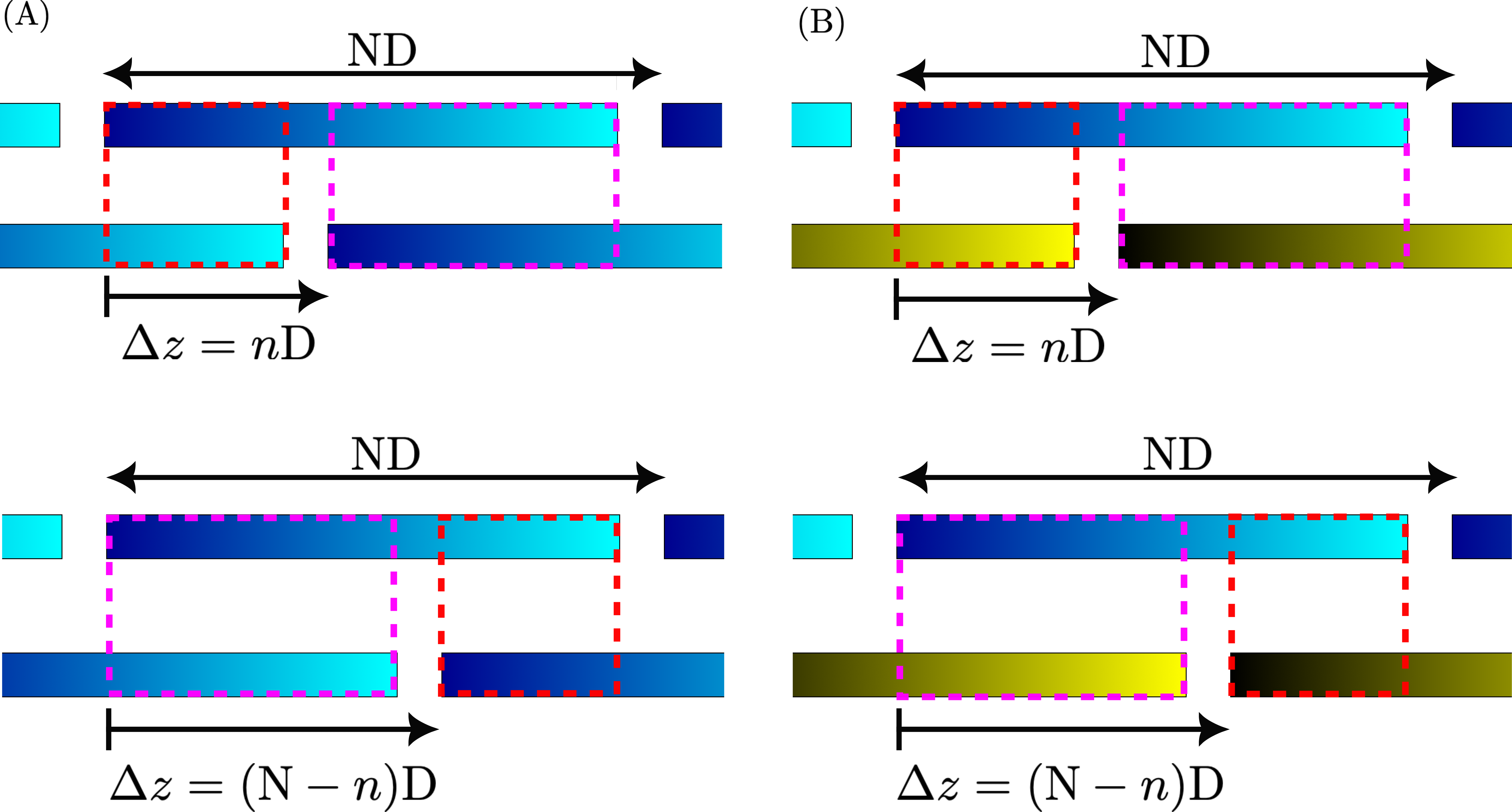}
\caption{Schematic representation of pairwise, ND axially periodic tropocollagen interactions showing: (A). Equivalence of interactions at $\Delta z = n\text{D}$ and $\Delta z = (\text{N}-n)\text{D}$ for each molecule contributing an identical set of residues. (B). Non-equivalence of interactions at $\Delta z = n\text{D}$ and $\Delta z = (\text{N}-n)\text{D}$ when each molecule contributes a distinct set of residues. Different colours represent distinct residue contributions.}
\label{Stagger_Symmetry}
\end{figure}

\clearpage
\bibliography{sample}

\begin{thebibliography}{56}
\providecommand{\natexlab}[1]{#1}
\providecommand{\url}[1]{\texttt{#1}}
\expandafter\ifx\csname urlstyle\endcsname\relax
  \providecommand{\doi}[1]{doi: #1}\else
  \providecommand{\doi}{doi: \begingroup \urlstyle{rm}\Url}\fi

\bibitem[Bella and Hulmes(2017)]{bella2017fibrillar}
Jordi Bella and David~JS Hulmes.
\newblock "fibrillar collagens".
\newblock In \emph{Fibrous proteins: structures and mechanisms}, pages 457--490. Springer, 2017.

\bibitem[Reznikov et~al.(2018)Reznikov, Bilton, Lari, Stevens, and Kr{\"o}ger]{reznikov2018fractal}
Natalie Reznikov, Matthew Bilton, Leonardo Lari, Molly~M Stevens, and Roland Kr{\"o}ger.
\newblock Fractal-like hierarchical organization of bone begins at the nanoscale.
\newblock \emph{Science}, 360\penalty0 (6388):\penalty0 eaao2189, 2018.

\bibitem[Koorman et~al.(2022)Koorman, Jansen, Khalil, Haughton, Visser, R{\"a}tze, Haakma, Sakalauskait{\`e}, van Diest, de~Rooij, et~al.]{koorman2022spatial}
Thijs Koorman, Karin~A Jansen, Antoine Khalil, Peter~D Haughton, Daan Visser, Max~AK R{\"a}tze, Wisse~E Haakma, Gabriel{\`e} Sakalauskait{\`e}, Paul~J van Diest, Johan de~Rooij, et~al.
\newblock Spatial collagen stiffening promotes collective breast cancer cell invasion by reinforcing extracellular matrix alignment.
\newblock \emph{Oncogene}, 41\penalty0 (17):\penalty0 2458--2469, 2022.

\bibitem[Towe(1970)]{towe1970oxygen}
Kenneth~M Towe.
\newblock Oxygen-collagen priority and the early metazoan fossil record.
\newblock \emph{Proceedings of the National Academy of Sciences}, 65\penalty0 (4):\penalty0 781--788, 1970.

\bibitem[Fidler et~al.(2017)Fidler, Darris, Chetyrkin, Pedchenko, Boudko, Brown, Gray~Jerome, Hudson, Rokas, and Hudson]{fidler2017collagen}
Aaron~L Fidler, Carl~E Darris, Sergei~V Chetyrkin, Vadim~K Pedchenko, Sergei~P Boudko, Kyle~L Brown, W~Gray~Jerome, Julie~K Hudson, Antonis Rokas, and Billy~G Hudson.
\newblock Collagen iv and basement membrane at the evolutionary dawn of metazoan tissues.
\newblock \emph{eLife}, 6:\penalty0 e24176, 2017.

\bibitem[O'Leary et~al.(2016)O'Leary, Wright, Brister, Ciufo, Haddad, McVeigh, Rajput, Robbertse, Smith-White, Ako-Adjei, et~al.]{o2016reference}
Nuala~A O'Leary, Mathew~W Wright, J~Rodney Brister, Stacy Ciufo, Diana Haddad, Rich McVeigh, Bhanu Rajput, Barbara Robbertse, Brian Smith-White, Danso Ako-Adjei, et~al.
\newblock Reference sequence (refseq) database at ncbi: current status, taxonomic expansion, and functional annotation.
\newblock \emph{Nucleic Acids Research}, 44\penalty0 (D1):\penalty0 D733--D745, 2016.

\bibitem[Yu et~al.(2011)Yu, Li, and Kim]{yu2011collagen}
S~Michael Yu, Yang Li, and Daniel Kim.
\newblock Collagen mimetic peptides: progress towards functional applications.
\newblock \emph{Soft Matter}, 7\penalty0 (18):\penalty0 7927--7938, 2011.

\bibitem[Xu and Kirchner(2021)]{xu2021collagen}
Yujia Xu and Michele Kirchner.
\newblock Collagen mimetic peptides.
\newblock \emph{Bioengineering}, 8\penalty0 (1):\penalty0 5, 2021.

\bibitem[Rezaei et~al.(2018)Rezaei, Lyons, and Forde]{rezaei2018environmentally}
Nagmeh Rezaei, Aaron Lyons, and Nancy~R Forde.
\newblock Environmentally controlled curvature of single collagen proteins.
\newblock \emph{Biophysical Journal}, 115\penalty0 (8):\penalty0 1457--1469, 2018.

\bibitem[Orgel et~al.(2006)Orgel, Irving, Miller, and Wess]{orgel2006microfibrillar}
Joseph~PRO Orgel, Thomas~C Irving, Andrew Miller, and Tim~J Wess.
\newblock Microfibrillar structure of type i collagen in situ.
\newblock \emph{Proceedings of the National Academy of Sciences}, 103\penalty0 (24):\penalty0 9001--9005, 2006.

\bibitem[Wieczorek et~al.(2015)Wieczorek, Rezaei, Chan, Xu, Panwar, Br{\"o}mme, Merschrod~S, and Forde]{wieczorek2015development}
Andrew Wieczorek, Naghmeh Rezaei, Clara~K Chan, Chuan Xu, Preety Panwar, Dieter Br{\"o}mme, Erika~F Merschrod~S, and Nancy~R Forde.
\newblock Development and characterization of a eukaryotic expression system for human type ii procollagen.
\newblock \emph{BMC biotechnology}, 15:\penalty0 1--17, 2015.

\bibitem[Revell et~al.(2021)Revell, Jensen, Shearer, Lu, Holmes, and Kadler]{revell2021collagen}
Christopher~K Revell, Oliver~E Jensen, Tom Shearer, Yinhui Lu, David~F Holmes, and Karl~E Kadler.
\newblock Collagen fibril assembly: New approaches to unanswered questions.
\newblock \emph{Matrix Biology Plus}, 12:\penalty0 100079, 2021.

\bibitem[Fang et~al.(2012)Fang, Goldstein, Turner, Les, Orr, Fisher, Welch, Rothman, and Banaszak~Holl]{fang2012type}
Ming Fang, Elizabeth~L Goldstein, A~Simon Turner, Clifford~M Les, Bradford~G Orr, Gary~J Fisher, Kathleen~B Welch, Edward~D Rothman, and Mark~M Banaszak~Holl.
\newblock Type i collagen d-spacing in fibril bundles of dermis, tendon, and bone: bridging between nano-and micro-level tissue hierarchy.
\newblock \emph{ACS nano}, 6\penalty0 (11):\penalty0 9503--9514, 2012.

\bibitem[Chen et~al.(2017)Chen, Ahn, Col{\'o}n-Bernal, Kim, and Banaszak~Holl]{chen2017relationship}
Junjie Chen, Taeyong Ahn, Isabel~D Col{\'o}n-Bernal, Jinhee Kim, and Mark~M Banaszak~Holl.
\newblock The relationship of collagen structural and compositional heterogeneity to tissue mechanical properties: a chemical perspective.
\newblock \emph{ACS nano}, 11\penalty0 (11):\penalty0 10665--10671, 2017.

\bibitem[Petruska and Hodge(1964)]{petruska1964subunit}
John~A Petruska and Alan~J Hodge.
\newblock A subunit model for the tropocollagen macromolecule.
\newblock \emph{Proceedings of the National Academy of Sciences}, 51\penalty0 (5):\penalty0 871--876, 1964.

\bibitem[Smith(1968)]{smith1968molecular}
JW~Smith.
\newblock Molecular pattern in native collagen.
\newblock \emph{Nature}, 219\penalty0 (5150):\penalty0 157--158, 1968.

\bibitem[Trus and Piez(1980)]{trus1980compressed}
Benes~L Trus and Karl~A Piez.
\newblock Compressed microfibril models of the native collagen fibril.
\newblock \emph{Nature}, 286\penalty0 (5770):\penalty0 300--301, 1980.

\bibitem[Orgel et~al.(2001)Orgel, Miller, Irving, Fischetti, Hammersley, and Wess]{orgel2001situ}
Joseph~PRO Orgel, Andrew Miller, Thomas~C Irving, Robert~F Fischetti, Andrew~P Hammersley, and Tim~J Wess.
\newblock The in situ supermolecular structure of type i collagen.
\newblock \emph{Structure}, 9\penalty0 (11):\penalty0 1061--1069, 2001.

\bibitem[Hulmes et~al.(1973)Hulmes, Miller, Parry, Piez, and Woodhead-Galloway]{hulmes1973analysis}
David~JS Hulmes, Andrew Miller, David~AD Parry, Karl~A Piez, and John Woodhead-Galloway.
\newblock Analysis of the primary structure of collagen for the origins of molecular packing.
\newblock \emph{Journal of Molecular Biology}, 79\penalty0 (1):\penalty0 137--148, 1973.

\bibitem[Trus and Piez(1976)]{trus1976molecular}
Benes~L Trus and Karl~A Piez.
\newblock Molecular packing of collagen: three-dimensional analysis of electrostatic interactions.
\newblock \emph{Journal of Molecular Biology}, 108\penalty0 (4):\penalty0 705--732, 1976.

\bibitem[Piez and Trus(1978)]{piez1978sequence}
Karl~A Piez and Benes~L Trus.
\newblock Sequence regularities and packing of collagen molecules.
\newblock \emph{Journal of Molecular Biology}, 122\penalty0 (4):\penalty0 419--432, 1978.

\bibitem[Hofmann et~al.(1978)Hofmann, Fietzek, and K{\"u}hn]{hofmann1978role}
H~Hofmann, PP~Fietzek, and K~K{\"u}hn.
\newblock The role of polar and hydrophobic interactions for the molecular packing of type i collagen: a three-dimensional evaluation of the amino acid sequence.
\newblock \emph{Journal of Molecular Biology}, 125\penalty0 (2):\penalty0 137--165, 1978.

\bibitem[Chen et~al.(1991)Chen, Kung, Feairheller, and Brown]{chen1991energetic}
James~M Chen, Chun~E Kung, Stephen~H Feairheller, and Eleanor~M Brown.
\newblock An energetic evaluation of a “smith” collagen microfibril model.
\newblock \emph{Journal of Protein Chemistry}, 10:\penalty0 535--552, 1991.

\bibitem[Puszkarska et~al.(2022)Puszkarska, Frenkel, Colwell, and Duer]{puszkarska2022using}
Anna~M Puszkarska, Daan Frenkel, Lucy~J Colwell, and Melinda~J Duer.
\newblock Using sequence data to predict the self-assembly of supramolecular collagen structures.
\newblock \emph{Biophysical Journal}, 121\penalty0 (16):\penalty0 3023--3033, 2022.

\bibitem[Harris and Lewis(2016)]{harris2016collagen}
J~Robin Harris and Richard~J Lewis.
\newblock The collagen type i segment long spacing (sls) and fibrillar forms: Formation by atp and sulphonated diazo dyes.
\newblock \emph{Micron}, 86:\penalty0 36--47, 2016.

\bibitem[Hulmes et~al.(1983)Hulmes, Bruns, and Gross]{hulmes1983state}
DJ~Hulmes, Romaine~R Bruns, and Jerome Gross.
\newblock On the state of aggregation of newly secreted procollagen.
\newblock \emph{Proceedings of the National Academy of Sciences}, 80\penalty0 (2):\penalty0 388--392, 1983.

\bibitem[Stylianou(2022)]{stylianou2022assessing}
Andreas Stylianou.
\newblock Assessing collagen d-band periodicity with atomic force microscopy.
\newblock \emph{Materials}, 15\penalty0 (4):\penalty0 1608, 2022.

\bibitem[Bella(2010)]{bella2010new}
Jordi Bella.
\newblock A new method for describing the helical conformation of collagen: Dependence of the triple helical twist on amino acid sequence.
\newblock \emph{Journal of Structural Biology}, 170\penalty0 (2):\penalty0 377--391, 2010.

\bibitem[Rainey and Goh(2002)]{rainey2002statistically}
Jan~K Rainey and M~Cynthia Goh.
\newblock A statistically derived parameterization for the collagen triple-helix.
\newblock \emph{Protein Science}, 11\penalty0 (11):\penalty0 2748--2754, 2002.

\bibitem[Orgel et~al.(2014)Orgel, Persikov, and Antipova]{orgel2014variation}
Joseph~PRO Orgel, Anton~V Persikov, and Olga Antipova.
\newblock Variation in the helical structure of native collagen.
\newblock \emph{PLoS One}, 9\penalty0 (2):\penalty0 e89519, 2014.

\bibitem[Neukirch et~al.(2008)Neukirch, Goriely, and Hausrath]{neukirch2008chirality}
S{\'e}bastien Neukirch, Alain Goriely, and Andrew~C Hausrath.
\newblock Chirality of coiled coils: elasticity matters.
\newblock \emph{Physical Review Letters}, 100\penalty0 (3):\penalty0 038105, 2008.

\bibitem[Liu et~al.(2004)Liu, Yong, Deng, Kallenbach, and Lu]{liu2004atomic}
Jie Liu, Wei Yong, Yiqun Deng, Neville~R Kallenbach, and Min Lu.
\newblock Atomic structure of a tryptophan-zipper pentamer.
\newblock \emph{Proceedings of the National Academy of Sciences}, 101\penalty0 (46):\penalty0 16156--16161, 2004.

\bibitem[Baselt et~al.(1993)Baselt, Revel, and Baldeschwieler]{baselt1993subfibrillar}
David~R Baselt, Jean-Paul Revel, and John~D Baldeschwieler.
\newblock Subfibrillar structure of type i collagen observed by atomic force microscopy.
\newblock \emph{Biophysical Journal}, 65\penalty0 (6):\penalty0 2644--2655, 1993.

\bibitem[Raspanti et~al.(2018)Raspanti, Reguzzoni, Protasoni, and Basso]{raspanti2018not}
Mario Raspanti, Marcella Reguzzoni, Marina Protasoni, and Petra Basso.
\newblock Not only tendons: The other architecture of collagen fibrils.
\newblock \emph{International Journal of Biological Macromolecules}, 107:\penalty0 1668--1674, 2018.

\bibitem[Mason and Arndt(2004)]{mason2004coiled}
Jody~M Mason and Katja~M Arndt.
\newblock Coiled coil domains: stability, specificity, and biological implications.
\newblock \emph{ChemBioChem}, 5\penalty0 (2):\penalty0 170--176, 2004.

\bibitem[Fraser and TP(1973)]{fraser1973conformation}
RDBM Fraser and MacRae TP.
\newblock \emph{Conformation in fibrous proteins and related synthetic polypeptides}.
\newblock New York: Academic Press, 1973.

\bibitem[McBride~Jr et~al.(1997)McBride~Jr, Choe, Shapiro, and Brodsky]{mcbride1997altered}
Daniel~J McBride~Jr, Vincent Choe, Jay~R Shapiro, and Barbara Brodsky.
\newblock Altered collagen structure in mouse tail tendon lacking the $\alpha$2 (i) chain.
\newblock \emph{Journal of Molecular Biology}, 270\penalty0 (2):\penalty0 275--284, 1997.

\bibitem[Antipova and Orgel(2010)]{antipova2010situ}
Olga Antipova and Joseph~PRO Orgel.
\newblock In situ d-periodic molecular structure of type ii collagen.
\newblock \emph{Journal of Biological Chemistry}, 285\penalty0 (10):\penalty0 7087--7096, 2010.

\bibitem[Asgari et~al.(2017)Asgari, Latifi, Heris, Vali, and Mongeau]{asgari2017vitro}
Meisam Asgari, Neda Latifi, Hossein~K Heris, Hojatollah Vali, and Luc Mongeau.
\newblock In vitro fibrillogenesis of tropocollagen type iii in collagen type i affects its relative fibrillar topology and mechanics.
\newblock \emph{Scientific Reports}, 7\penalty0 (1):\penalty0 1392, 2017.

\bibitem[Brodsky et~al.(1980)Brodsky, Eikenberry, and Cassidy]{brodsky1980unusual}
Barbara Brodsky, Eric~F Eikenberry, and Kathleen Cassidy.
\newblock An unusual collagen periodicity in skin.
\newblock \emph{Biochimica et Biophysica Acta (BBA)-Protein Structure}, 621\penalty0 (1):\penalty0 162--166, 1980.

\bibitem[Chanut-Delalande et~al.(2001)Chanut-Delalande, Fichard, Bernocco, Garrone, Hulmes, and Ruggiero]{chanut2001control}
H{\'e}lene Chanut-Delalande, Agnes Fichard, Simonetta Bernocco, Robert Garrone, David~JS Hulmes, and Florence Ruggiero.
\newblock Control of heterotypic fibril formation by collagen v is determined by chain stoichiometry.
\newblock \emph{Journal of Biological Chemistry}, 276\penalty0 (26):\penalty0 24352--24359, 2001.

\bibitem[Mizuno et~al.(2013)Mizuno, B{\"a}chinger, Imamura, Hayashi, and Adachi]{mizuno2013fragility}
Kazunori Mizuno, Hans~Peter B{\"a}chinger, Yasutada Imamura, Toshihiko Hayashi, and Eijiro Adachi.
\newblock Fragility of reconstituted type v collagen fibrils with the chain composition of $\alpha$1 (v) $\alpha$2 (v) $\alpha$3 (v) respective of the d-periodic banding pattern.
\newblock \emph{Connective Tissue Research}, 54\penalty0 (1):\penalty0 41--48, 2013.

\bibitem[Hansen and Bruckner(2003)]{hansen2003macromolecular}
Uwe Hansen and Peter Bruckner.
\newblock Macromolecular specificity of collagen fibrillogenesis: fibrils of collagens i and xi contain a heterotypic alloyed core and a collagen i sheath.
\newblock \emph{Journal of Biological Chemistry}, 278\penalty0 (39):\penalty0 37352--37359, 2003.

\bibitem[Plumb et~al.(2007)Plumb, Dhir, Mironov, Ferrara, Poulsom, Kadler, Thornton, Briggs, and Boot-Handford]{plumb2007collagen}
Darren~A Plumb, Vivek Dhir, Aleksandr Mironov, Laila Ferrara, Richard Poulsom, Karl~E Kadler, David~J Thornton, Michael~D Briggs, and Raymond~P Boot-Handford.
\newblock Collagen xxvii is developmentally regulated and forms thin fibrillar structures distinct from those of classical vertebrate fibrillar collagens.
\newblock \emph{Journal of Biological Chemistry}, 282\penalty0 (17):\penalty0 12791--12795, 2007.

\bibitem[Leikina et~al.(2002)Leikina, Mertts, Kuznetsova, and Leikin]{leikina2002type}
E~Leikina, MV~Mertts, N~Kuznetsova, and S~Leikin.
\newblock Type i collagen is thermally unstable at body temperature.
\newblock \emph{Proceedings of the National Academy of Sciences}, 99\penalty0 (3):\penalty0 1314--1318, 2002.

\bibitem[Xu and Kirchner(2022)]{xu2022segment}
Yujia Xu and Michele Kirchner.
\newblock "segment-long-spacing (sls) and the polymorphic structures of fibrillar collagen".
\newblock In \emph{Macromolecular Protein Complexes IV: Structure and Function}, pages 495--521. Springer, 2022.

\bibitem[Perret et~al.(2001)Perret, Merle, Bernocco, Berland, Garrone, Hulmes, Theisen, and Ruggiero]{perret2001unhydroxylated}
St{\'e}phanie Perret, Christine Merle, Simonetta Bernocco, Patricia Berland, Robert Garrone, David~JS Hulmes, Manfred Theisen, and Florence Ruggiero.
\newblock Unhydroxylated triple helical collagen i produced in transgenic plants provides new clues on the role of hydroxyproline in collagen folding and fibril formation.
\newblock \emph{Journal of Biological Chemistry}, 276\penalty0 (47):\penalty0 43693--43698, 2001.

\bibitem[Hwang et~al.(2010)Hwang, Thiagarajan, Parmar, and Brodsky]{hwang2010interruptions}
Eileen~S Hwang, Geetha Thiagarajan, Avanish~S Parmar, and Barbara Brodsky.
\newblock Interruptions in the collagen repeating tripeptide pattern can promote supramolecular association.
\newblock \emph{Protein Science}, 19\penalty0 (5):\penalty0 1053--1064, 2010.

\bibitem[Boot-Handford et~al.(2003)Boot-Handford, Tuckwell, Plumb, Rock, and Poulsom]{boot2003novel}
Raymond~P Boot-Handford, Danny~S Tuckwell, Darren~A Plumb, Claire~Farrington Rock, and Richard Poulsom.
\newblock A novel and highly conserved collagen (pro$\alpha$1 (xxvii)) with a unique expression pattern and unusual molecular characteristics establishes a new clade within the vertebrate fibrillar collagen family.
\newblock \emph{Journal of Biological Chemistry}, 278\penalty0 (33):\penalty0 31067--31077, 2003.

\bibitem[Koch et~al.(2003)Koch, Laub, Zhou, Hahn, Tanaka, Burgeson, Gerecke, Ramirez, and Gordon]{koch2003collagen}
Manuel Koch, Friedrich Laub, Peihong Zhou, Rita~A Hahn, Shizuko Tanaka, Robert~E Burgeson, Donald~R Gerecke, Francesco Ramirez, and Marion~K Gordon.
\newblock Collagen xxiv, a vertebrate fibrillar collagen with structural features of invertebrate collagens: selective expression in developing cornea and bone.
\newblock \emph{Journal of Biological Chemistry}, 278\penalty0 (44):\penalty0 43236--43244, 2003.

\bibitem[Forlino et~al.(2011)Forlino, Cabral, Barnes, and Marini]{forlino2011osteogenImperfecta}
Antonella Forlino, Wayne~A Cabral, Aileen~M Barnes, and Joan~C Marini.
\newblock New perspectives on osteogenesis imperfecta.
\newblock \emph{Nature Reviews Endocrinology}, 7\penalty0 (9):\penalty0 540--557, 2011.

\bibitem[Venturoni et~al.(2003)Venturoni, Gutsmann, Fantner, Kindt, and Hansma]{venturoni2003investigations}
Manuela Venturoni, Thomas Gutsmann, Georg~E Fantner, Johannes~H Kindt, and Paul~K Hansma.
\newblock Investigations into the polymorphism of rat tail tendon fibrils using atomic force microscopy.
\newblock \emph{Biochemical and Biophysical Research Communications}, 303\penalty0 (2):\penalty0 508--513, 2003.

\bibitem[Chen et~al.(2019)Chen, Strawn, and Xu]{Chen2019-ra}
Fangfang Chen, Rebecca Strawn, and Yujia Xu.
\newblock The predominant roles of the sequence periodicity in the self-assembly of collagen-mimetic mini-fibrils.
\newblock \emph{Protein Sci.}, 28\penalty0 (9):\penalty0 1640--1651, September 2019.

\bibitem[Kuznetsova and Leikin(1999)]{kuznetsova1999does}
Natalia Kuznetsova and Sergey Leikin.
\newblock Does the triple helical domain of type i collagen encode molecular recognition and fiber assembly while telopeptides serve as catalytic domains?: effect of proteolytic cleavage on fibrillogenesis and on collagen-collagen interaction in fibers.
\newblock \emph{Journal of Biological Chemistry}, 274\penalty0 (51):\penalty0 36083--36088, 1999.

\bibitem[Kawashima and Kanehisa(2000)]{kawashima2000aaindex}
Shuichi Kawashima and Minoru Kanehisa.
\newblock Aaindex: amino acid index database.
\newblock \emph{Nucleic Acids Research}, 28\penalty0 (1):\penalty0 374--374, 2000.

\bibitem[De~Bruijn(2014)]{de2014asymptotic}
Nicolaas~Govert De~Bruijn.
\newblock \emph{Asymptotic methods in analysis}.
\newblock Courier Corporation, 2014.

\end{thebibliography}

\end{document}